\def\year{2026}\relax
\documentclass[letterpaper]{article} 
\usepackage{aaai22}  
\usepackage{times}  
\usepackage{helvet}  
\usepackage{courier}  
\usepackage[hyphens]{url}  
\usepackage{graphicx} 
\usepackage{natbib}  
\usepackage{caption} 
\DeclareCaptionStyle{ruled}{labelfont=normalfont,labelsep=colon,strut=off} 
\usepackage{algorithm}
\usepackage{algorithmic}
\usepackage{amsmath}

\usepackage{booktabs}
\usepackage{multirow}
\usepackage{array}

\usepackage{tabularx}
\usepackage{longtable}

\usepackage[dvipsnames]{xcolor}

\usepackage{subfigure} 

\usepackage{newfloat}
\usepackage{listings}
\floatstyle{ruled}
\newfloat{listing}{tb}{lst}{}
\floatname{listing}{Listing}
\title{\textbf{From Attention to Dialogue: Does Audience Engagement Reinforce Constructive Cross-Party Communication?}}

\author{
    Ahana Biswas, 
    Yu-Ru Lin
}
\date{}
\affiliations{
    University of Pittsburgh\\
    
    ahana.biswas@pitt.edu, 
    yurulin@pitt.edu
}

\usepackage{booktabs}
\usepackage{threeparttable}

\usepackage{soul}
\usepackage{xspace}

\def \ahb #1{{\color{black}{#1} }} 
\def \ah #1{{\color{black}{#1} }} 
\def \ahf #1{{\color{black}{#1} }}
\def \ahff #1{{\color{black}{#1} }}
\def \ab #1{{\color{black}{#1} }}

\newcommand{\yrv}[1]{{\textcolor{black}{#1}}}

\newcommand{\yrvv}[1]{{\textcolor{black}{#1}}} 

\newcommand{\RBox}{\raisebox{0.7ex}{\colorbox{BrickRed}{\hspace{0.1em}\vspace{0.5ex}}}\xspace}
\newcommand{\DBox}{\raisebox{0.7ex}{\colorbox{RoyalBlue}{\hspace{0.1em}\vspace{0.5ex}}}\xspace}
\newcommand{\RtoD}{R$\rightarrow$D\xspace}
\newcommand{\DtoR}{D$\rightarrow$R\xspace}

\usepackage{tikz}
\newcommand*\circled[1]{\tikz[baseline=(char.base))]{
            \node[shape=circle,draw,inner sep=.5pt] (char) {\scriptsize #1};}}

\def \a #1{{\color{black}{#1} }}
\def \x #1{{\color{black}{#1} }}

\def \b #1{{\color{black}{#1}\xspace}}

\begin{document}

\maketitle

\begin{abstract}


\x{While existing works have emphasized how elites shape mass opinion, we ask whether the reverse also holds: do audience reactions on social media actively shape elite behavior?}\a{We examine this question through the lens of cross-partisan interactions (CPIs), which can either foster deliberation or deepen polarization. Using a dataset of over 1.1 million cross-party retweets, replies, and mentions between U.S. state legislators and their audiences on Twitter/X (2020–2021), we first establish baseline patterns of \b{engagement}: Democrats gain modest \b{engagement} in replies and mentions, while Republicans often face penalties in direct cross-party interactions. Building on this, we show that audience engagement produces a feedback loop that conditions future elite behavior. Following highly visible CPIs, legislators are not only more likely to engage again in cross-talk, but also shift their rhetorical strategies. \b{Engagement} consistently promotes causal reasoning, subjective language, and positive-emotion framing in subsequent CPIs. \b{These findings suggest a positive association between audience engagement and constructive cross-party discourse among elites, challenging overly simplified interpretations in the literature that emphasize social media as a primary driver of rising or falling polarization.}}

\end{abstract}

\section{Introduction}

\a{Cross-partisan interactions (CPIs)—instances of engagement across party lines—are often regarded as vital to democratic discourse. Such exchanges can foster deliberation and reduce stereotyping across divides~\cite{habermas1989structural,pettigrew2006meta}. Yet they can also trigger backlash, amplify hostility, and heighten affective polarization~\cite{mutz2002consequences,bail2018exposure}. This tension raises a fundamental question: do CPIs help depolarize, or do they reinforce partisan divides?}

\a{This question is especially pressing in the case of political elites, whose communication carries disproportionate weight in shaping public opinion. Elite cue theory holds that political leaders provide informational shortcuts that shape how citizens interpret and respond to political issues~\cite{zaller1992nature}. \x{More recently, research on online influencers highlights how highly visible figures strategically manage their self-presentation and audience engagement in order to capture and sustain attention~\cite{tufekci2013not}.} Legislators today occupy both roles: as elites, their communication has downstream consequences for public opinion, and as visible figures on social media, they are embedded in dynamics where \b{engagement} is currency. This dual role raises an important question: \textit{In the attention economy, do audiences shape elites’ behavior by rewarding certain cross-party interactions while discouraging others?}}

\begin{figure}[ht]
\centering
\includegraphics[width=\linewidth]{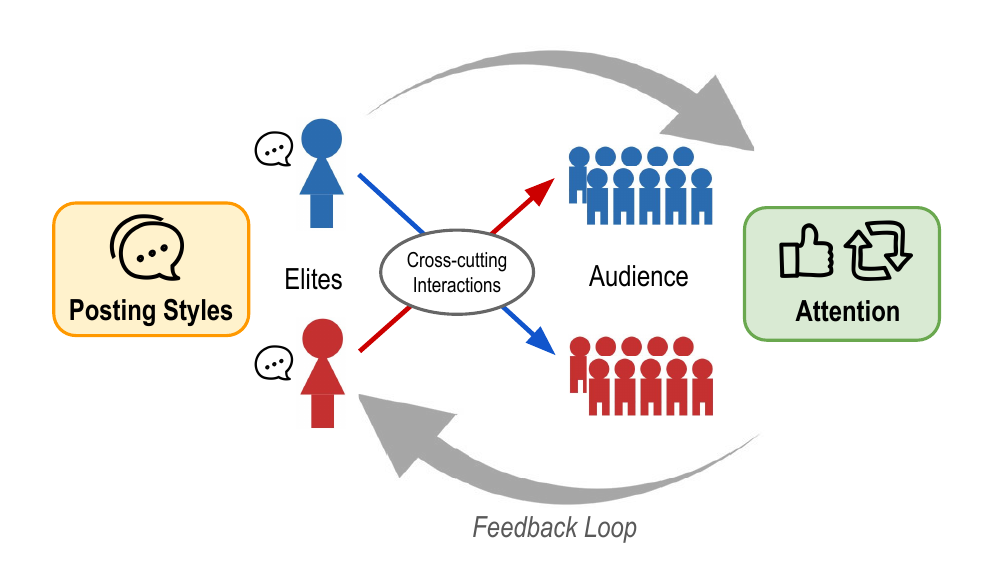}
\caption{{\bf Study Design.} 
\a{We analyze how cross-party interactions (CPI) shape the online \b{engagement} of political elites and how \b{engagement} feeds back into future behavior. We test whether CPIs boost or reduce \b{engagement}, and which posting styles drive these effects (RQ1), and how \b{engagement} shapes both the frequency (RQ2) and style (RQ3) of future CPI.}
} \label{fig:vis}
\vspace{-1.5em}
\end{figure}

\a{To investigate this possibility, we deliberately narrow our scope. First, we focus on political elites rather than general users, since their communication is uniquely positioned to shape democratic discourse. Legislators, while not influencers in the conventional sense, serve as a critical proxy for elites whose online behavior has both symbolic and strategic weight. Second, we examine cross-partisan posts, as these interactions are especially consequential for polarization and depolarization~\cite{nyhan2010corrections,garimella2018political}. Third, we concentrate on audience\footnote{\b{We use the term \emph{audience} to refer to users who engage with a legislator’s posts. For instance, references to “Republican (Democratic) audiences” denote the audiences of Republican (Democratic) legislators.}} engagement, since \b{engagement} is the primary way in which publics reward or penalize elites online and prior research suggests it conditions future communicative behavior~\cite{slater2007reinforcing}. Finally, we study not only whether elites engage across party lines, but also how they do so—capturing both the frequency and stylistic choices of CPIs as outcomes that may shift in response to prior \b{engagement}\footnote{\b{Throughout the paper, we use the term {\it engagement} to denote a user-normalized measure of audience response, defined relative to each legislator's own recent posting history. Specifically, engagement captures whether a given post receives more or fewer responses than that legislator's typical recent posts, rather than measuring absolute exposure or impressions (see Section 3.3).}}.}


\a{To conceptualize these dynamics, we adopt the stimulus–organism–response (SOR) model~\cite{jacoby2002stimulus}. In this framework, audience engagement functions as the stimulus, elites’ interpretation of these signals represents the organismal process, and subsequent posting frequency and style are the responses. This lens allows us to move beyond static assessments of \b{engagement} to test whether attention feeds back into elites’ future communication strategies. Our focus thus shifts from audience reactions alone to the feedback loops that connect audience engagement and elite behavior—an aspect of bipartisan communication that has received limited attention in prior research. \x{In this work, we focus on the positive direction of this loop, asking how increases in the \b{engagement} of CPIs shape the likelihood and style of future cross-talk. We hypothesize that positive reinforcement dominates because platforms foreground visible signals of success (e.g., {\it likes, replies, shares}), giving elites stronger incentives to repeat attention-getting behaviors.}}

Our study leverages over 1.2M interactions between U.S. state legislators and their audiences on Twitter/X during 2020–2021. We organize our analysis around four RQs:
\\




    


\noindent \textbf{RQ1:} \x{How does cross-partisan communication gain or lose \b{engagement} in the attention economy?}
    \begin{itemize}
        \item \textbf{RQ1a:} \x{To what extent do cross-party interactions, compared to intra-party ones, shape the \b{engagement} of elites?}
        \item \textbf{RQ1b:} \x{How does the nature of cross-talk (e.g., tone, framing, topics) affect their \b{engagement}?}
    \end{itemize}

\noindent \textbf{RQ2:} \a{How does the \b{engagement} of past cross-party interactions affect how often legislators engage in future cross-partisan communication?}
    
\noindent \textbf{RQ3:} \a{To what extent does the \b{engagement} of past cross-party interactions affect the rhetorical or stylistic choices legislators make in future cross-partisan communication?}
\\

\a{RQ1a establishes the baseline \b{engagement} effects of CPIs, RQ1b examines which posting styles explain these effects, and together they set the stage for RQ2 and RQ3, which test whether audience responses feed back into the frequency and style of elites’ future cross-partisan behavior (Fig. \ref{fig:vis}\footnote{Fig. 1 icons by Andrew Aprilio Raharjo (CC BY 3.0)}).}

\a{By framing CPIs as part of a dynamic feedback loop, this study extends prior work on cross-cutting exposure and partisan discourse by centering elite behavioral change. Our findings offer new insight into whether social media incentivizes constructive cross-party dialogue or reinforces divisive, attention-maximizing strategies—with direct implications for understanding polarization and the health of democratic communication.}

\paragraph{Findings.} 
\a{Our findings reveal marked asymmetries. Republican legislators see lower \b{engagement} when engaging via direct modes of communication like replies and mentions, whereas Democrats experience modest \b{engagement} gains (+4.5\% and +8\%). Content also matters. Incivility in cross-party replies increases Republican \b{engagement} by 18\%, while uncivil mentions reduce \b{engagement} for both parties. Importantly, higher engagement with CPIs is linked to increased bipartisan communication in subsequent weeks: both parties increase mentions (by 2.2\% for Republicans, 1.3\% for Democrats) following highly visible cross-party posts. Beyond frequency, we also find that \b{engagement} conditions how legislators communicate in future CPIs. For replies, \b{engagement} \b{is positively associated with} hedged (+0.9\%) and subjective (+1.2\%) framing for Democrats, and subjective and causal framing (both by 0.9\%) for Republicans. For mentions, Democrats lean into subjective and positive-emotion expression by 0.7\%), while Republicans reduce URLs (-1.5\%) but amplify hedging, causality, positive, and issue-centered mentions by around 1\%. Taken together, these results provide evidence of a \b{engagement-driven} feedback loop in which audience responses shape not only whether but also how elites engage across party lines.}

\paragraph{Implications.}
\a{Prior work on digital political communication often highlights how audience engagement reinforces polarizing behaviors, such as antagonism or incivility, raising concerns about the incentives built into attention-driven platforms (Mutz, 2002; Bail et al., 2018; Cinelli et al., 2021). 
\b{Our findings extend this literature by showing that engagement is associated not only with amplification or polarization, but also with subsequent shifts in communicative style. Specifically, higher engagement in CPIs is positively associated with increased use of subjective, causal, and topical language in later replies and mentions. These stylistic features are commonly linked to explanation, contextualization, and issue-focused discussion, which are often considered components of deliberative exchange. These results suggest that audience engagement may coincide with communicative adjustments in cross-party interaction. This points to a more nuanced role for engagement in democratic communication: under some conditions, engagement may be aligned with discourse strategies that support more substantive interaction across partisan lines, rather than uniformly reinforcing polarization.}}

\section{Related Work}

\a{\subsubsection{Exposure to Cross-Cutting Discussions}
Scholars have long debated whether cross-cutting exposure strengthens or weakens democracy. Some view it as fostering deliberation and reducing stereotyping across divides~\cite{habermas1989structural,barber2003strong,pettigrew2006meta}, while others argue it discourages participation or exacerbates polarization~\cite{mutz2002consequences,taber2006motivated,nyhan2010corrections}. Empirical work documents both dynamics: disagreeable exposure can mobilize political participation~\cite{lu2016cross,kim2016social,min2018all}, but can also provoke backfire effects~\cite{nyhan2010corrections,taber2006motivated}. These effects are often asymmetric across the ideological spectrum, as liberals and conservatives prioritize distinct values~\cite{graham2009liberals,jost2007needs,grossmann2016asymmetric}. Computational studies extend this debate to online contexts: heterogeneous discussions do occur~\cite{an2019political,wu2021cross}, yet they can also intensify divides~\cite{bail2018exposure}. Moreover, bipartisan actors are often less central and attract lower engagement~\cite{garimella2018political}. Taken together, this body of work underscores that engagement with CPIs is far from uniform. Some contexts promote deliberation and broaden participation, while others intensify polarization. These mixed findings raise a critical question about how engagement dynamics contribute to cross-talk online.}

\a{\subsubsection{Nature and Style of Political Cross-Talk}
Cross-party communication is a recurring but contested feature of online discourse~\cite{wu2021cross,de2021no}. Some interactions aim at genuine deliberation, while others resemble trolling or provocation~\cite{phillips2015we,flores2018mobilizing,hua2020towards}. Linguistic and emotional patterns also vary: users adapt their style when addressing out-partisans compared to co-partisans~\cite{an2019political}, and CPIs are often more negative or emotional~\cite{burrell2020would,wu2021cross,marchal2022nice}. Broader studies show partisan divides in framing during political crises~\cite{demszky2019analyzing}, yet linguistic cues such as positivity or reduced subjectivity can foster engagement across ideological lines~\cite{saveski2022engaging}. These highlight the importance of studying the role of linguistic features and topics in driving content engagement, as they can influence how CPIs resonate across ideological divides and either foster more inclusive or divisive political discourse.}

\a{\subsubsection{Feedback Loops and Elite Adaptation}
Political science has long emphasized that elites are responsive to public signals and strategically adapt their communication~\cite{downs1957economic,carrubba2001electoral,gabel2007estimating,santoro2021exploring}. Traditionally, responsiveness was inferred from elections, opinion polls, or media coverage. On social media, elites instead receive instantaneous feedback in the form of engagement metrics that signal audience approval. Computational studies highlight how such feedback loops shape behavior in digital environments: scientists adapt their messaging after viral attention~\cite{hasan2022impact}, newsrooms shift coverage toward high-engagement stories~\cite{caplan2016controls}, and creators adjust strategies to algorithmic incentives~\cite{bishop2019managing}. Feedback also shapes political behavior, as algorithmic amplification can reinforce user actions including misinformation sharing~\cite{cinelli2021echo,hanley2023sub}. Communication research has long theorized reinforcing spirals between exposure and expression~\cite{slater2007reinforcing}, but large-scale empirical evidence of feedback-driven adaptation in elite cross-party discourse remains limited. Whether audience engagement systematically shapes elites’ CPIs is an open question.}

\a{\paragraph{Present Work.}
Despite extensive work on cross-cutting exposure, cross-talk styles, and elite responsiveness, few studies examine whether audience engagement feeds back into elite cross-party behavior. Prior work shows that bipartisan actors often struggle for \b{engagement}~\cite{garimella2018political}, and that audiences are more responsive to out-party criticism than in-party praise~\cite{yu2024partisanship}. However, these studies do not examine if \b{engagement} systematically alters elite strategies over time.
We address this gap by analyzing over one million CPIs between U.S. state legislators and their audiences on Twitter/X. Our study proceeds in two stages. First, we examine how cross-party interactions affect elite \b{engagement} and how posting styles shape engagement outcomes \textbf{(RQ1)}. Second, we investigate whether \b{engagement} feeds back into elite behavior by influencing the frequency \textbf{(RQ2)} and posting style \textbf{(RQ3)} of future CPIs.
By linking audience engagement to elite adaptation, we position CPIs as embedded in a feedback system, where \b{engagement} acts as a reinforcement signal.}

\section{\x{Study Design}}\label{sec:sd}


\b{Our study proceeds in two stages to examine how \b{engagement} shapes cross-partisan interactions. First, we analyze how different forms of cross-party engagement and associated rhetorical and stylistic choices resonate with audiences in the attention economy. Building on these patterns of audience response, we then examine how engagement signals feedback into elites' subsequent communicative behavior. The following sections describe the data, variables, and modeling strategy used in our analysis.}

\subsection{Dataset}\label{sec:data}

\yrv{We use the Twitter dataset provided by~\citet{biswas2025political}, which contains the complete posting history of all U.S. state representatives and senators who held office at any point during 2020-2021. This dataset includes approximately 4 million tweets posted by 3,568 (38.8\%) U.S. state legislators, of which 53.4\% are Democrats, responsible for 71.6\% of the posts, and 45.3\% are Republicans, contributing 28.3\% of the posts\footnote{The number of Independent legislators (N=29) is too small for meaningful analysis and has thus been excluded from the study.}. \b{Out of the these, around 2.3M (57.5\%) tweets had atleast one interaction (i.e., either replied to/mentioned/retweeted another user)\footnote{\b{This accounts for interactions with 1.7M unique users. Around 73\% of these interactions originate from Democrat legislators.}}. }
Following prior work~\cite{biswas2025political}, we measure the engagement with each post by summing the number of likes, retweets, replies, and quotes.}

\yrv{Additionally, we collect all available follower accounts from a 2021 data snapshot, which allows us to construct a follower network of the legislators. This network consists of approximately 3.8 million connections and over 1.3 million unique follower accounts. The set of followers ($U_F$) and the set of legislators ($U_L$) together form the set of users ($U = U_F \cup U_L$) considered in this study.}

\yrv{\noindent\paragraph{Cross-cutting interactions.} To explore the relationship between political interactions and public engagement, we define ``{cross-cutting}'' interactions as communication that occurs across partisan lines. 
\yrvv{Specifically, we define a legislator's post as {\it cross-cutting} if it mentions, replies to, or retweets a user in $U$---whether a legislator or follower---whose political leaning opposes their own\footnote{Interactions are included in our dataset if a legislator engages with a user's content (e.g., retweets, mentions, or replies) and the user's political leaning can be inferred from the follower network, even if the user does not follow the legislator.}.}
While the political affiliations of legislators are readily available, the leanings of their followers are not. To infer the political leaning of a follower $f \in U_F$, let $D_f$ and $R_f$ represent the number of Democratic and Republican legislators, respectively, that user $f$ follows. We include only those followers who follow at least three legislators to ensure a more reliable estimation of their leaning.}

\yrv{The leaning score $l_f$ of each user $f$ is calculated as the normalized difference between the number of Democratic ($D_f$) and Republican ($R_f$) legislators followed by the user. \ab{We refer to this metric as the ``attention leaning'' since it reflects the partisan distribution of legislator accounts followed.} Specifically, the score is given by:}

\begin{figure}[ht]
\centering
\includegraphics[width=0.75\linewidth]{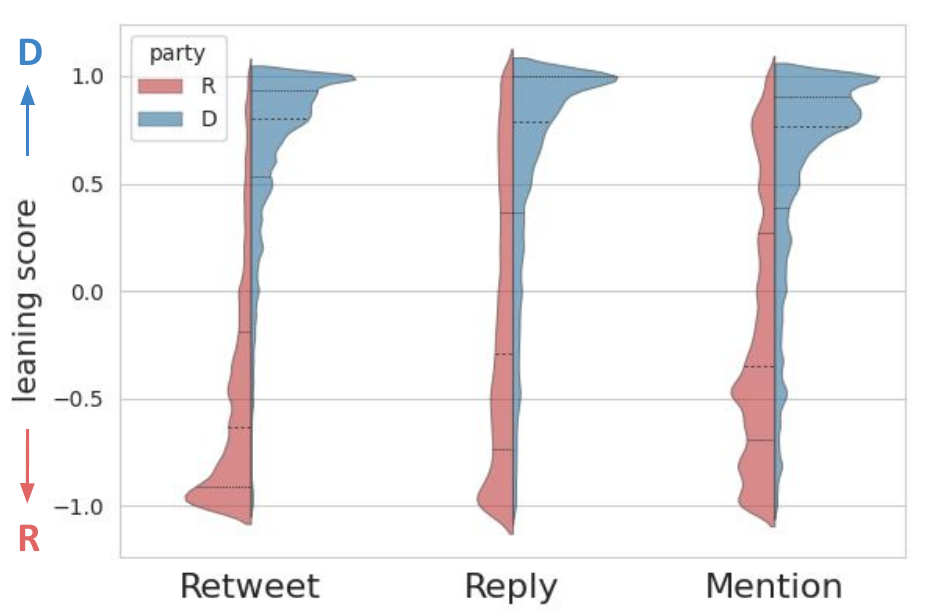}
\vspace{-0.5em}
\caption{{\bf Distribution of \ab{attention}scores.} Legislators from both parties primarily engage with \ahf{users}having the same political leaning across all interaction types. Republicans tend to engage in more cross-talks than Democrats.} \label{fig:lean}
\vspace{-0.5em}
\end{figure}

\vspace{-0.5em}
\begin{equation}
    l_f = \frac{D_f-R_f}{D_f + R_f}
\vspace{-.2em}
\end{equation}
\yrv{This results in a score ranging from -1 to 1, where a score closer to 1 indicates a stronger alignment with Democrats, and a score closer to -1 indicates a stronger alignment with Republicans. Users who follow an equal number of Democrat and Republican legislators will have a score of 0, indicating no clear partisan preference.
Figure~\ref{fig:lean} shows the distribution of \ab{attention} leaning scores for various interaction types~\footnote{\ab{Mentions in our analysis do not account for replies, i.e., an account is considered ``mentioned'' only if it explicitly appears in the text body of the post as a mention.}}. This distribution highlights the political polarization in the interactions, where Democrat legislators predominantly interact with users whose scores are closer to 1, while Republicans tend to interact with users whose scores are closer to -1, suggesting that legislators' interactions on Twitter are heavily aligned with users who share their political affiliation.} 

\yrv{Based on the distribution of \ab{attention}leaning scores and manual annotation of a sample of their Twitter accounts (see Appendix for details), we set the threshold for identifying Democrat-leaning followers at a score of 0.71 or higher and Republican-leaning followers at -0.69 or lower\ahff{\footnote{\ahff{This yields over 53.6K (3\% of all users interacted with) unique users with either Republican or Democrat leaning. \b{These users appear to be highly politically attentive: they follow multiple legislators and account for a disproportionate share (approximately 50\%) of elite-audience interactions.}}}.}}\ahb{Interactions with users from the same political leaning are considered as {\it intra-partisan} posting.} \b{Using our method we are able to identify the partisan leaning of users interacted with for around 1.17M (50.8\%) posts by 3,287 (92.1\%)\footnote{\b{On average, we are able to estimate 55\% interaction directions (i.e., cross-party or intra-party) for each legislator overall (49\% and 56\% for Republicans and Democrats respectively).}} legislators which is the final dataset used in our study. Among these around 82K (7\%) posts have atleast one cross-cutting interaction\footnote{\b{Notably, a single tweet can contain multiple interactions.}}.}

\begin{table}[ht]

\setlength{\tabcolsep}{3pt}
\renewcommand{\arraystretch}{1.3}
\centering

\caption{\b{Volume of cross-cutting posts by each party and their median engagement. The numbers in the brackets denote the ratio of not cross-cutting to cross-cutting posts.}}\label{tab:data_stats}

\begin{tabular}{cccc}
\toprule
 &  & N & Engagement \\
\midrule
\multirow{2}{*}{Retweet} & R $\rightarrow$ D & 13,520 (10.9) & 29.0 (6.4) \\
                            & D $\rightarrow$ R & 8,595 (60.9) & 3291.5 (0.007) \\
\midrule
\multirow{2}{*}{Reply} & R $\rightarrow$ D & 11,133 (2.1) & 1.0 (1.0) \\
                            & D $\rightarrow$ R & 7,040 (16.8) & 1.0 (2.0) \\
\midrule
\multirow{2}{*}{Mention} & R $\rightarrow$ D & 35,709 (2.2) & 5.0 (1.2) \\
                            & D $\rightarrow$ R &  29,168 (17.9) & 8.0 (0.75) \\
\bottomrule
\end{tabular}
\end{table}


\b{Table \ref{tab:data_stats} shows the volume and}
\ahb{Figure~\ref{fig:data} \ab{in Appendix}shows the \b{rate} of cross-cutting and intra-partisan posts and their engagement. \ah{Republican legislators have 5 to 6 times more cross-partisan interaction rates than their Democratic counterparts (Fig.~\ref{fig:data}A). For instance, among all replies from Republican legislators, 32\% are cross-partisan, compared to only 5.6\% of replies from Democrats.}
\a{The engagement metric for retweets returned by the Twitter API
inherently reflect the popularity of the original post (see Appendix), therefore we only report results for mention and reply in the paper. Results for retweet are provided in the Appendix.}
\yrv{The typical engagements on cross-cutting posts are usually lower than intra-partisan ones for Republicans and in the case of replies for Democrats.}
The engagement is notably higher when Democrats retweet users having Republican leaning (Fig.~\ref{fig:data}B).} 







\subsection{\ahb{Individuals’ attributes}} \label{sec:att}
\yrv{We characterize legislators by their activities on the platform and individual-level characteristics. To measure platform activity and influence, we include features such as post count, follower count, and in-degree centrality within the follower network, serving as proxies for their network influence~\cite{hasan2022impact}.}

\ahb{The individual-level attributes include socio-demographic features such as state, gender (Men vs. Women), ethnicity\footnote{Ethnicity and gender are mapped using Ballotpedia. Binary genders used due to lack of sufficient data on non-binary genders.} (White vs. Non-White), and ideology scores \ab{\cite{biswas2025political}.} The dataset consists of approximately 68.2\% men and 62.2\% White legislators. We leverage the ideology scores constructed by \citet{shor2011ideological} where the scores range between -2.3 (extremely liberal) to +2.3 (extremely conservative).} 

\ahb{In addition to the socio-demographics, we include two other individual-level attributes, (i) central (C) vs. local (L) and (ii) audience partisanship.}

\paragraph{\bf Central (C) vs. local (L).}

We categorize legislators as central, local, or balanced based on the geographic diversity of their audiences. Central legislators, with broader cross-state followings, can shape national discourse by engaging conversations that span diverse viewpoints~\cite{mutz2001facilitating,jun2014political}. Their CPI may help disrupt echo chambers by connecting opposing perspectives. Local legislators, by contrast, may concentrate on state-specific audiences, tailoring messages to regional concerns and making cross-partisan communication more salient at the local level. Audience locality is measured as the fraction of followers from a legislator’s home state; higher scores indicate more localized audiences (See Appendix). Legislators in the top and bottom 30 percentiles are classified as local and central, respectively.



\paragraph{\bf \b{Audience Partisanship}.} 
\ahb{CPIs can be effective in bridging political divides or may have backfire effects depending on the partisan composition of a legislator’s audience.} 
\b{Audience partisanship captures the extent to which a legislator’s audience leans toward one political party, which can shape how cross-partisan interactions are received.}
\b{We operationalize audience partisanship using the attention-leaning score of followers~\cite{garimella2017long}. Specifically, the audience partisanship score of a legislator is measured as the median leaning score of users who engage with that legislator’s posts.}
\b{This measure does not capture ideological homogeneity or affective polarization within the audience, but rather provides a summary measure of partisan orientation that is relevant for understanding how CPIs are evaluated by different party audiences.}

\subsection{\ab{\b{Measuring Engagement}}}\label{sec:vis}
\ab{Our measure of \b{engagement}, termed as {\it overperforming} score, is drawn from prior work by \citet{biswas2025political}\footnote{This score is inspired by the metric used at Meta CrowdTangle and generalized to be suitable for other platforms like Twitter. The engagement components (likes, shares, replies, and quotes) used to get {\it \b{engagement}} are similarly used on Twitter (https://github.com/twitter/the-algorithm?tab=readme-ov-file) to calculate a post's expected engagement.
}.  
The overperforming score for post $i$ is,
$y_{ui} = v_{ui} / (b_{u} + b_{0})$,
where $v_{ui}$ is the engagement received on post $i$, $b_u$ is the median engagement for legislator $u$'s posts on the platform in previous $m$-days and a threshold ($b_{0}$) for the minimum number of engagements on a post to be considered as overperforming. We choose the $b_{0}=10$ and $m=14$ as suggested by \citet{biswas2025political} \ab{(See Appendix)}. 
\b{This metric captures engagement relative to a legislator’s own recent baseline, thereby normalizing for stable differences across legislators (e.g., audience size) and slowly varying platform-level factors that affect engagement broadly. While we do not explicitly model platform ranking algorithms or exposure, the baseline normalization reduces sensitivity to global shifts in platform dynamics that influence engagement across posts within the same time window.}}

\section{Methods}
\yrv{In this section, we begin by outlining the characteristics of legislators' posting styles and discussing their relevance. Finally, we detail the methods used to address our research questions.}

\subsection{Characterizing Posting Styles}\label{sec:ps}

\a{We select posting styles commonly studied in political communication. Broadly, we study two categories: rhetorical styles, which capture how legislators frame arguments and reasoning (e.g., causality terms, generalization, hedging, topic indicators, URL usage), and stylistic features, which reflect tone and affect (e.g., incivility, positive or negative emotion, subjectivity). This allows us to examine how different dimensions of communication map onto audience engagement and future cross-party interactions.}

\subsubsection{Linguistic markers.} 
\yrv{We use four linguistic cues---{\it emotion}, {\it subjectivity}, {\it generalization}, and {\it argumentation}---to characterize the legislators' posting styles.} \ab{These features capture distinct aspects of political communication---emotion highlights the intensity of posts, subjectivity reflects the degree of personal opinion, generalization assesses accessibility and inclusiveness, and argumentation measures logical reasoning and persuasive quality.}
Below we describe our reasoning and the choice of measures for these markers.

\textbf{Emotion.} Prior work shows that emotionally charged content elicits stronger audience responses and engagement~\cite{mutz2007effects}. Emotional language can enhance the perceived sincerity of a message, shaping both attention and public opinion. In polarized contexts, such appeals may reinforce or challenge existing beliefs, making emotion a key marker for CPI. We measure emotions using positive and negative word lists from Linguistic Inquiry and Word Count (LIWC)\footnote{https://www.liwc.app/}. Among negative emotions, we focus on anger and anxiety, which are especially influential in political discourse~\cite{de2011two,ryan2012makes}.

\textbf{Subjectivity.} Subjectivity refers to the expression of opinions and evaluations~\cite{wiebe2000learning}. In cross-party interactions, subjective language can increase engagement by presenting relatable or persuasive viewpoints, but may also polarize by emphasizing partisan perspectives. We measure subjectivity using OpinionFinder’s Subjectivity lexicon~\cite{wilson2005opinionfinder}.

\textbf{Generalization.} Generalizations can make cross-party messages more accessible to broad audiences but also risk stereotyping and reducing nuance, potentially reinforcing divides. Analyzing generalizations helps assess whether such communication balances simplicity with accuracy and inclusiveness in shaping engagement. We measure them using MPQA’s\footnote{\ab{https://mpqa.cs.pitt.edu/lexicons/arg\_lexicon/}}generalization lexicon.




\textbf{Argumentation.} Argumentation conveys reasoning and persuasion, making it central to constructive CPIs. Well-structured arguments can foster productive communication and engage diverse audiences. We capture argumentation through \textit{hedges} and \textit{causation}\footnote{\ahb{Argumentation mining is not feasible due to short text length.}}. Hedge words signal uncertainty and can soften confrontational tones, fostering receptiveness across divides, and are measured using the lexicon from~\citet{islam2020lexicon}. Causation terms clarify connections between events and provide logical structure, enhancing persuasiveness in cross-party dialogue; we use word lists from MPQA and LIWC to measure them.


\ab{We evaluate the quality of the lexicon-based markers by manually labeling around 40 samples per linguistic feature (Appendix). The precision of certain markers differs across the parties, possibly due to the distinct communication styles employed by the parties. To mitigate this variation, our study design incorporates a comparative approach by evaluating cross-party interactions relative to intra-party interactions as described in later sections. This design helps to isolate the effects of cross-party communication while minimizing biases arising from party-specific linguistic styles. \b{Additionally, the prevalence of each posting style and representative examples for both parties are reported in Appendix Table~\ref{tab:style_prevalence_examples}.}}

\subsubsection{Harmful content.} {\it Harmful} content, including uncivil language, and untrustworthy or non-credible information can play a significant role in shaping the dynamics of cross-party discussions, especially in politically charged environments \ab{\cite{goovaerts2020uncivil,hanley2023sub}.} 

\textbf{Incivility.} Incivility refers to disrespectful, hostile, or derogatory language. In cross-party interactions, it can escalate tensions, polarize audiences, and discourage constructive communication. We assess incivility using toxicity scores, a common practice in prior work~\cite{frimer2023incivility,kim2021distorting}. Following~\cite{biswas2025political}, we use Detoxify to generate toxicity scores and apply a threshold of 0.82\footnote{\citet{biswas2025political} identified 0.82 as the optimal threshold for this dataset based on manual labeling}, classifying posts above this cutoff as uncivil. Around 2.3\% of Democratic CPIs (0.6\% intra-partisan) and 1.1\% of Republican CPIs (0.7\% intra-partisan) are uncivil.

\textbf{Low-credibility.} Low-credibility content in cross-party discussions can undermine debate quality, reduce trust, and mislead the public. When political elites share such information, they risk perpetuating false narratives and weakening democratic processes. We detect low-credibility content using labels from~\citet{tai2023official}, which rely on Media Bias/Fact Check (MBFC) credibility ratings of URL domains\footnote{\b{Domain-level information was extracted from the expanded URL metadata when available, or from the visible domain portion of shortened or truncated URLs, which is sufficient for identifying source domains.}}, a common practice in prior work~\cite{lasser2022social}. Around 0.3\% of CPIs from both Republicans (2.1\% intra-partisan) and Democrats (negligible intra-partisan) contain low-credibility information (see Appendix).


\subsubsection{Topics.} Topics frame the content of CPI and shape both the nature of interactions and public engagement. We classify posts using a keyword-based approach followed by semantic clustering, covering salient issues in U.S. politics such as \textit{BLM, COVID-19, rights, immigration, gun control, climate, abortion}, and the \textit{Capitol riots} (see Appendix). Posts without relevant keywords are labeled ``other,'' while those matching multiple categories are assigned to all relevant topics to capture complex, multi-faceted discussions. 


\subsubsection{External Information.} 
\yrv{Posts containing URLs may provide additional information, attracting users with diverse perspectives and encouraging engagement across viewpoints. We consider the inclusion of a URL\footnote{\ab{We do not examine URL types (e.g., finance, sports) in detail, which could offer additional insights but is limited by the challenge of recovering full URLs for all posts.}} as a feature of posting style that could influence the \b{engagement} of CPIs.}

\paragraph{Differences in cross-party vs. intra-party posting styles.}

\a{We compare posting styles\footnote{All posting styles are binary, i.e., 1 if present. For linguistic markers, the marker is considered present if the lexicons appear at least once.} and legislators’ attributes across cross- and intra-party interactions using Mann–Whitney U tests with Bonferroni correction. Effect sizes (rank-biserial correlations) are reported in Appendix Fig.\ref{fig:mwu} (confidence intervals in Fig.\ref{fig:rq0_ci}).}

\a{Clear differences emerge. Negative emotions such as anger, anxiety, and incivility appear more often in deliberative forms of cross-talk (replies and mentions), echoing prior findings that cross-partisan communication tends to be more toxic or antagonistic~\cite{burrell2020would,wu2021cross}. Legislator attributes also matter: among Democrats, central legislators are more likely to engage across party lines, while among Republicans, local legislators take the lead. In both parties, legislators with polarized audiences engage less in all forms of cross-talk, and those at the ideological extremes are least likely to participate, consistent with theories of selective engagement~\cite{heatherly2017filtering}.}

\a{These results suggest that bipartisan communication is shaped not only by rhetorical styles but also by legislators’ representational contexts and audience structures. These variations can strongly influence the \b{engagement} of CPIs and are incorporated into our analyses in the following sections.}

\subsection{Measuring the Effect of Cross-cutting Interactions on \b{Engagement} (RQ1)}
We estimate the observed causal effect of cross-cutting interactions on \b{engagement} using a two-fold approach: matching and regression analysis.

\subsubsection{Matching.} \a{CPIs systematically differ from intra-party ones in ways that may also affect \b{engagement}. To reduce this bias, we match each CPI to an intra-party counterpart within the same interaction type (retweet, reply, mention) and party. This addresses potential confounders such as topical focus (e.g., gun control), rhetoric strategies, legislator attributes (Section~\ref{sec:ps}), and timing (e.g., election periods). Posts are represented using RoBERTa embeddings~\cite{liu2019roberta}, concatenated with legislator attributes and timing; for low-credibility content, URL headlines are also included. Posts shorter than 10 words are excluded. Following prior work~\cite{hasan2022impact,sahly2019social,biswas2025political}, these covariates capture key textual and contextual factors. We then perform 1:1 K-Nearest Neighbors matching, which achieves covariate balance (Appendix Fig.~\ref{Fig:2}).} 

\subsubsection{Regression Analysis (RQ1a).} Using the matched posts, we estimate the \ab{(observed)} effect of cross-cutting interactions using a linear mixed effects model,

\vspace{-1em}
\begin{equation}\label{rq1_eq}
    y_{ui} \sim \alpha_0 + \alpha_1\Theta_{ui} + \vec{\alpha}_XX_u + r_u + d_i
\end{equation}

where $y_{ui}$ is the overperforming score of legislator $u$'s post $i$ and $\Theta_{ui}$ is a binary variable indicating whether the post is cross-cutting (i.e., $\Theta=1$). Therefore, $\alpha_1$ captures the effect of cross-cutting interactions on the \b{engagement} of legislators' posts. $X_u$ is the vector of individual attributes described in Section~\ref{sec:att} used to control for confounding at the user-level. We incorporate random effects on the user and post timing (i.e., day count) denoted by $r_u$ and $d_i$ respectively---to control for unobserved heterogeneity, i.e., factors that vary across users or time periods but are not explicitly included in our model. This is done to reduce bias and improve the robustness of the estimated effects. 

\subsubsection{Effect of Posting Styles (RQ1b).}

\yrvv{We analyze how posting styles (see Section~\ref{sec:ps}) are associated with the \b{engagement} of cross-cutting posts. Unlike RQ1, which examines a single treatment variable, we decompose the variable from \ref{rq1_eq} into two parts: the effects of posting styles and other uncaptured factors.}
We use the following regression analysis,


\vspace{-1em}
\begin{equation}\label{rq2_eq}
    y_{ui} \sim \alpha_0 + \alpha_{L^-}\Theta_{ui} + \vec{\alpha}_LL_{ui}\Theta_{ui} + \vec{\alpha}_XX_u + r_u + d_i 
\end{equation}

where $L_{ui}$ is the vector of posting styles of legislator $u$’s post $i$ and the coefficient vector $\vec{\alpha}_L$ represents the effect of each posting style on cross-talk \b{engagement}. \ahb{The coefficient $\alpha_{L^-}$ accounts for the treatment effects not captured by the posting style variables.} The other variables are same as described in equation~\ref{rq1_eq}. We similarly use a linear mixed effects model to estimate equation~\ref{rq2_eq}.

\subsection{\yrvv{Assessing the Link Between \b{Engagement} and Future Cross-party Interactions (RQ2)}}

In RQ2, we investigate how the \b{engagement} on cross-cutting interactions \ab{is linked to} the likelihood of future cross-party talks from political elites. 
\yrv{More specifically, we estimate how changes in \b{engagement} \ab{are correlated with} the rate of cross-party interactions in the subsequent week. This is done using the following regression analysis:}

\vspace{-1em}
\begin{equation}\label{rq3_eq}
    \rho_{u,t+1} \sim \beta_0 + \beta_1\delta_{u,t} + \vec{\beta}_XX_u + r_u + w_t + \rho_{u,t}
\end{equation}

\yrv{Here, $\rho_{u,t}$ and $\rho_{u,t+1}$ represent the fraction of cross-party posts made by legislator $u$ in weeks $t$ and $t+1$, respectively. The term $\delta_{u,t} = \text{frac}(y_{u,t|c} > 1) - \text{frac}(y_{u,t|nc} > 1)$ measures the difference between the fraction of overperforming cross-cutting posts and non-cross-cutting posts ($nc$) in week $t$. A post is considered ``overperforming'' if its \b{engagement} metrics exceed a threshold of 1. Non-cross-cutting ($nc$) posts refer to intra-partisan posts or posts with no interactions \ab{or interactions with users whose leaning is not known (i.e., all posts other than cross-cutting ones).} The coefficient $\beta_1$ represents the effect of changes in \b{engagement} on the future rate of cross-party interactions. To account for variability across individuals and time, we include random effects for users ($r_u$) and weeks ($w_t$), capturing individual-specific and temporal fluctuations, respectively. The term $\rho_{u,t}$ accounts for the autoregressive effect, where the current cross-cutting posting rate influences future interactions.}



\subsection{\a{\b{Engagement} and the Posting Style of Future Cross-party Interactions (RQ3)}}


\a{In RQ3, we examine whether the \b{engagement} gained from cross-party interactions b{is associated with} the stylistic choices elites adopt in subsequent cross-party communication. We test whether legislators use of particular posting styles—such as incivility, emotional tone, or topical framing changes depending on the relative \b{engagement} of their cross-partisan posts. We model this relationship as follows:}

\vspace{-1em}
\begin{equation}\label{rq4_eq}
\rho^{L}_{u,t+1} \sim \beta_0 + \beta_L \delta_{u,t} + \vec{\beta}_X X_u + r_u + w_t + \rho^{L}_{u,t}
\end{equation}
\vspace{-0.5em}

\a{Here, $\rho^{L}_{u,t+1}$ represents the rate of cross-partisan posts containing style $L$ (e.g., incivility, topics, linguistic markers) made by legislator $u$ in week $t+1$, while $\rho^{L}_{u,t}$ denotes the corresponding rate in week $t$. The term $\rho^{L}_{u,t}$ controls for autoregressive dynamics, ensuring that baseline tendencies in style usage are accounted for when estimating the effect of \b{engagement}. The remaining variables are the same as RQ2.}

\a{Posting styles include both linguistic and affective features as described in section \ref{sec:ps}, except topics are represented as binary indicators denoting whether a given issue is present in a post, allowing us to assess the overall use of topical content in CPIs. The coefficient $\beta_L$ measures the effect of \b{engagement} on the adoption of posting style $L$ in subsequent cross-party communication. A positive coefficient indicates that legislators use of style $L$ \b{is positively associated with} \b{engagement} of prior CPIs. We perform p-value correction (Bonferroni) for each interaction type and party, to account for multiple comparisons.}


To satisfy assumptions of linear regression, all continuous variables are suitably transformed to be close to normal distributions and standardized for equations~\ref{rq1_eq}, \ref{rq2_eq}, \ref{rq3_eq} and \ref{rq4_eq}. Therefore, the coefficients should be interpreted as changes in standard deviation (SD) units.

\section{Results}
\yrvv{This section presents results for each RQ, reporting standardized effect sizes in the main text. Coefficients $\alpha$ and $\beta$ reflect changes in standard deviations. Non-standardized effect sizes (Figs. ~\ref{fig:rq1_ns}--\ref{fig:rq3_ns}) and qualitative examples (Table~\ref{tab:rq2_example}) in the Appendix provide additional percentage-based and contextual interpretation.}

\subsection{\ah{RQ1a: What is the \b{Observed} Effect of Cross-cutting Interactions on \b{Engagement}?}}

\begin{figure}[ht]
\centering
\includegraphics[width=0.65\linewidth]{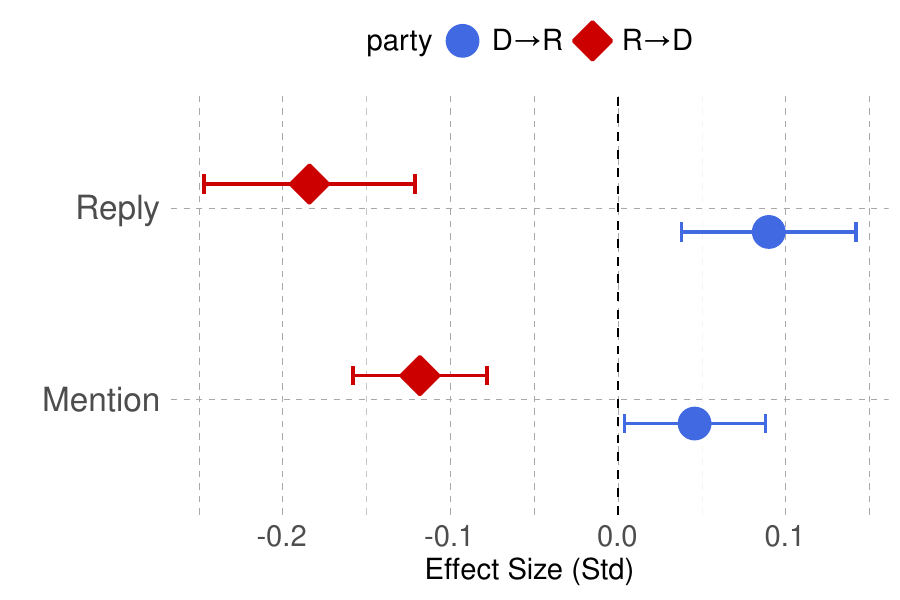}
\caption{{\bf Effect of CPIs on \b{engagement}.} \ab{\b{Engagement} of legislators varies significantly in cross-talks, with {\it asymmetric} effects depending on party affiliation and the nature of interactions. Effect size sign: (+) Higher (outperforming) \b{engagement} in CPI; (-) Lower (underperforming) \b{engagement} in CPI. Significant effects are denoted by filled markers ($p<0.05$).}} \label{fig:rq1}
\vspace{-1.5em}
\end{figure}

Figure~\ref{fig:rq1} shows the effect sizes and 95\% CI for the effect of cross-cutting interactions on \b{engagement}. 
When Republicans reply to ($\alpha_1$ = -0.18, $p<0.001$) or mention ($\alpha_1$ = -0.12, $p<0.001$) Democrats, their posts receive lower \b{engagement} compared to intra-party interactions. In contrast, Democrats gain higher \b{engagement} when replying to ($\alpha_1$ = 0.09, $p<0.001$) Republicans \ab{(corresponding to 8\% \b{engagement} gains as shown in Fig.~\ref{fig:rq1_ns})} or mentioning ($\alpha_1$ = 0.05, $p<0.05$) them.
This may indicate that Republican audiences are less receptive to cross-party engagements through replies or mentions, potentially leading to reduced \b{engagement}. Conversely, the positive effects for Democrats suggest their audiences may favor more intentional and direct cross-party interactions. 

\noindent \paragraph{Main Takeaways.} \ahb{Cross-cutting communication has an effect on the \b{engagement} of both Republican and Democrat legislators. \ah{Republican party audiences are less receptive to deliberative cross-party exchanges, while Democrat party audiences appear more interested in conversational engagements through mentions and replies, highlighting the asymmetries in political party audiences.} These differences could also arise from the nature of these communications which we further explore in RQ1b.} 



\subsection{\ah{RQ1b: What are the \b{Observed} Effects of Posting Styles on \b{Engagement} of CPIs?}}

\begin{figure*}[ht]
\centering
\includegraphics[width=0.75\linewidth]{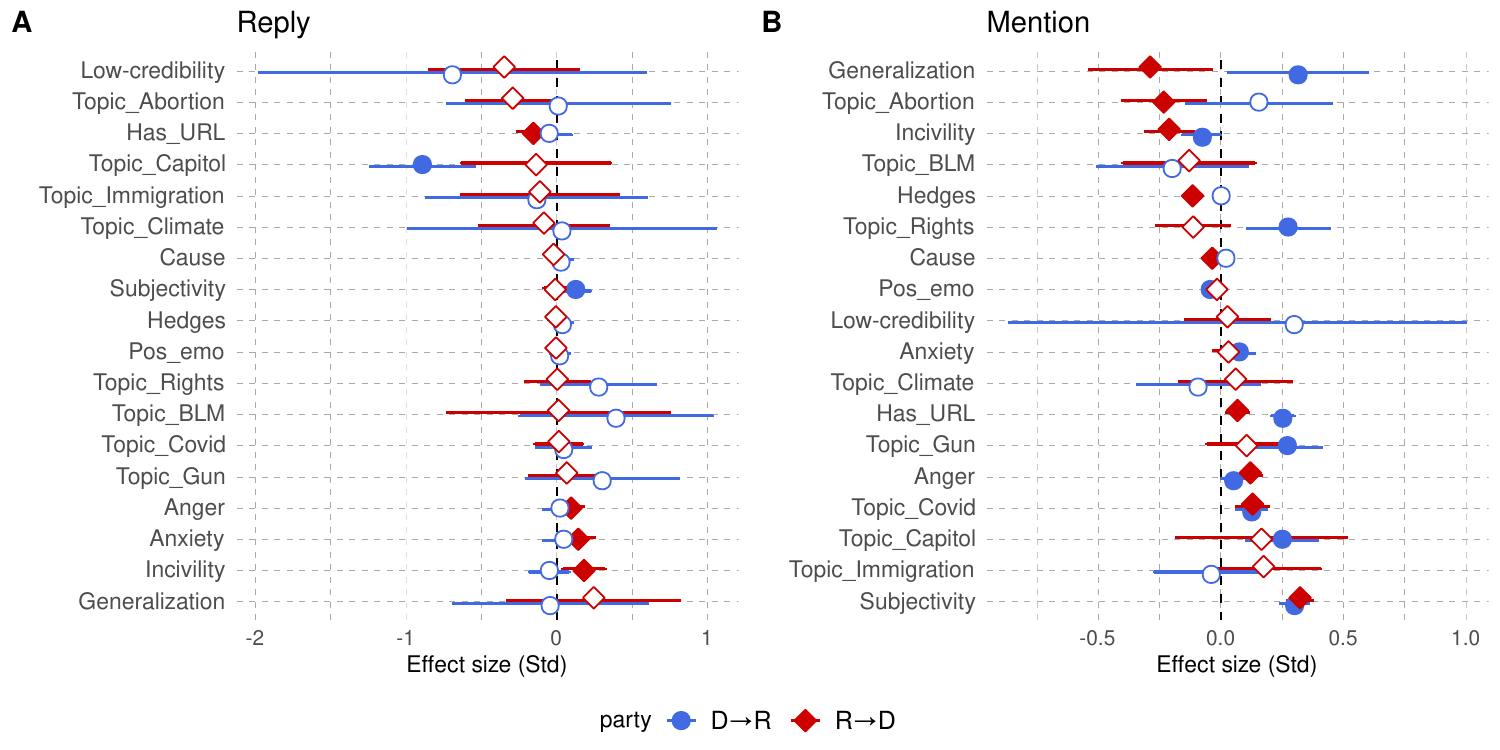}
\caption{{\bf Effect of posting styles on \b{engagement} of CPIs.} Posting styles are associated with the \b{engagement} garnered by CPIs and the associations differ across party and interaction types. \ab{ The covariates are sorted based on effect sizes for Republicans.}} \label{fig:rq2}
\vspace{-1.5em}
\end{figure*}



\a{Figure~\ref{fig:rq2} presents how different posting styles influence the \b{engagement} of CPIs. Details on significance levels appear in Appendix Table~\ref{tab:rq2_p}, with illustrative examples of posts provided in Appendix Table~\ref{tab:rq2_example}.}



\a{\paragraph{Reply.} In replies, \b{engagement} is strongly conditioned by both topic and tone. Democrats lose \b{engagement} when replying to Republicans about the Capitol riots ($\alpha_L=-0.89$). CPIs on such contentious topics, as shown in Appendix Table~\ref{tab:rq2_example} (\circled{11}, \circled{14}), often take the form of hostile rebuttals rather than substantive discussion, which likely discourages engagement. By contrast, Republicans gain \b{engagement} when their replies include incivility ($\alpha_L=0.18$) or negative emotions such as anger ($\alpha_L=0.09$) and anxiety ($\alpha_L=0.14$). Examples like \circled{15}, \circled{17} illustrate how emotionally charged replies can resonate with Republican audiences, even as they risk entrenching antagonistic rhetoric.}

\a{\paragraph{Mention.} In mentions, Republicans lose \b{engagement} when raising the topic of abortion ($\alpha_L=-0.23$), whereas Democrats gain attention when mentioning Republicans on rights ($\alpha_L=0.27$), gun control ($\alpha_L=0.27$), or the Capitol riots ($\alpha_L=0.25$). In Appendix Table~\ref{tab:rq2_example}, examples \circled{24} and \circled{26} illustrate how issue-centered mentions are rewarded, contrasting with combative replies. Both parties see \b{engagement} gains of about 13\% when mentioning opponents in the context of COVID-19 (Appendix Fig.~\ref{fig:rq2_ns}), suggesting bipartisan conversations on issues of national urgency resonate more broadly. Importantly, incivility in mentions reduces \b{engagement} for both Democrats (–8.3\%) and Republicans (–21\%), showing that audiences do not reward overt hostility in this interaction type. At the same time, the use of negative emotions modestly increases \b{engagement}: anger boosts engagement for both Democrats ($\alpha_L=0.07$) and Republicans ($\alpha_L=0.12$), while anxiety provides a boost for Democrats ($\alpha_L=0.05$).} 

\a{Other rhetorical markers also matter: URLs increase \b{engagement}, particularly for Democrats ($\alpha_L=0.25$ vs. $0.06$ for Republicans), while subjective language boosts engagement for both parties ($\alpha_L=0.30$ for Democrats; $0.32$ for Republicans). Hedging similarly lowers Republican \b{engagement} ($\alpha_L=-0.11$), suggesting their audiences respond unfavorably to uncertainty. Generalizations split audiences, enhancing Democratic mentions ($\alpha_L=0.31$) but reducing Republican ones ($\alpha_L=-0.28$). Examples \circled{29} and \circled{30} highlight these contrasting uses of generalization language.}

\noindent \paragraph{Main Takeaways.} 


\a{Posting styles are strongly \b{associated with} \b{engagement} of CPIs, with clear partisan asymmetries. In replies, Republican audiences reward antagonistic and emotional cues, while Democrats face \b{engagement} losses when engaging on contentious issues like Capitol riots. In mentions, Democratic audiences engage more with issue-centered discussions, URLs, and generalized framing, whereas Republicans are penalized for incivility, uncertainty, and generalization language. Both parties benefit when addressing nationally salient issues like COVID-19. Together, these \b{engagement} patterns provide the foundation for understanding how audience responses may shape the frequency and style of elites’ future CPI.}


\subsection{\ah{RQ2: What is the \b{Observed} Impact of \b{Engagement} on Future Cross-talks?}}

\begin{figure}[ht]
\centering
\includegraphics[width=0.65\linewidth]{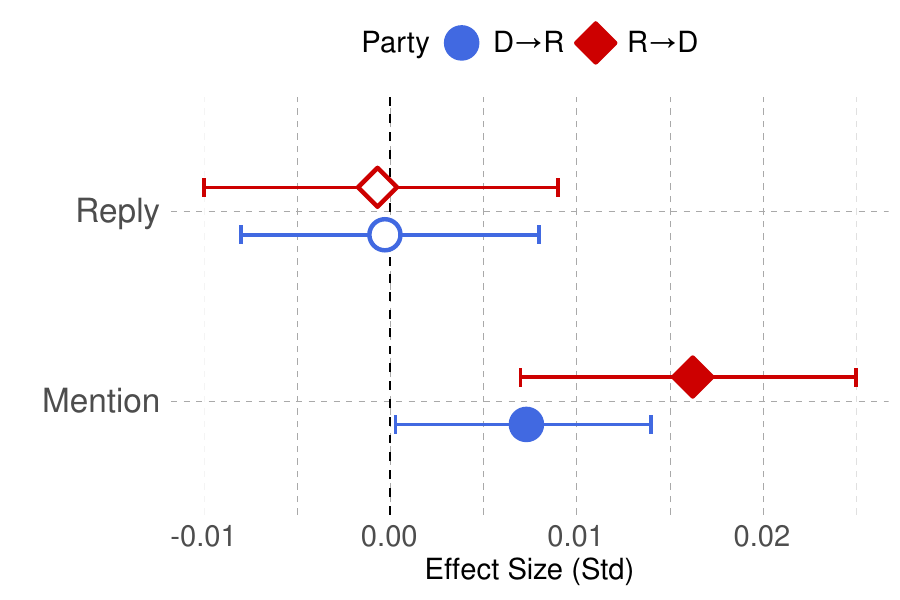}
\vspace{-0.5em}
\caption{\yrvv{{\bf Relationship between \b{engagement} and future CPIs.} Engagement with cross-partisan posts through mentions is strongly linked to increased future CPIs for both parties.}
} \label{fig:rq3}
\vspace{-1em}
\end{figure}

\yrvv{In RQ3, we examine how \b{engagement} from cross-cutting interactions is associated with the frequency of future cross-party communication by legislators.}
 Figure~\ref{fig:rq3} shows the effect sizes by type of interaction and party. 
For mentions, both Republicans ($\beta_1$ = 0.016, $p<0.001$) \ab{(corresponds to 2.1\% increase as shown in Fig.~\ref{fig:rq3_ns})} and Democrats ($\beta_1$ = 0.007, $p<0.05$) \ab{(1.3\% rise)} increase their cross-party mentions when these posts garner higher \b{engagement}. This suggests that legislators might recognize that engaging with the other side can amplify their reach and impact and hence tend to engage more in such cross-cutting discussions, more so in the case of Republicans. Though, in general, cross-party mentioning reduces the \b{engagement} of Republicans (RQ1), they tend to increase such interactions in the future when these garner higher attention, which highlights the subtleties of political communication online.

\yrv{We do not observe significant effects for cross-cutting replies. This suggests a feedback loop in which political elites tailor their content based on audience preferences, but this effect is limited to mentions. One explanation could be that replies receive less engagement than mentions, making the small \b{engagement} gains from replies insufficient to motivate political elites to prioritize such interactions in future conversations.}


\noindent \paragraph{Main Takeaways.}\a{Our findings point to a \b{engagement}-driven feedback loop in which audience reactions condition elite communication strategies. Legislators are more likely to sustain cross-party mentions after these interactions attract attention, suggesting that \b{engagement} acts as a reinforcement signal encouraging future bipartisan engagement. At the same time, the absence of effects for replies highlights that not all forms of dialogue generate the same incentives, underscoring the selective and complex ways in which online attention shapes elite cross-party behavior. Next, we examine whether \b{engagement} shapes not only the frequency of CPIs but also the rhetorical and stylistic choices elites adopt in future CPIs.}

\subsection{\a{RQ3: What is the \b{Observed} Impact of \b{Engagement} on the Style of Future Cross-talks?}}

\begin{figure*}[ht]
\centering
\includegraphics[width=0.75\linewidth]{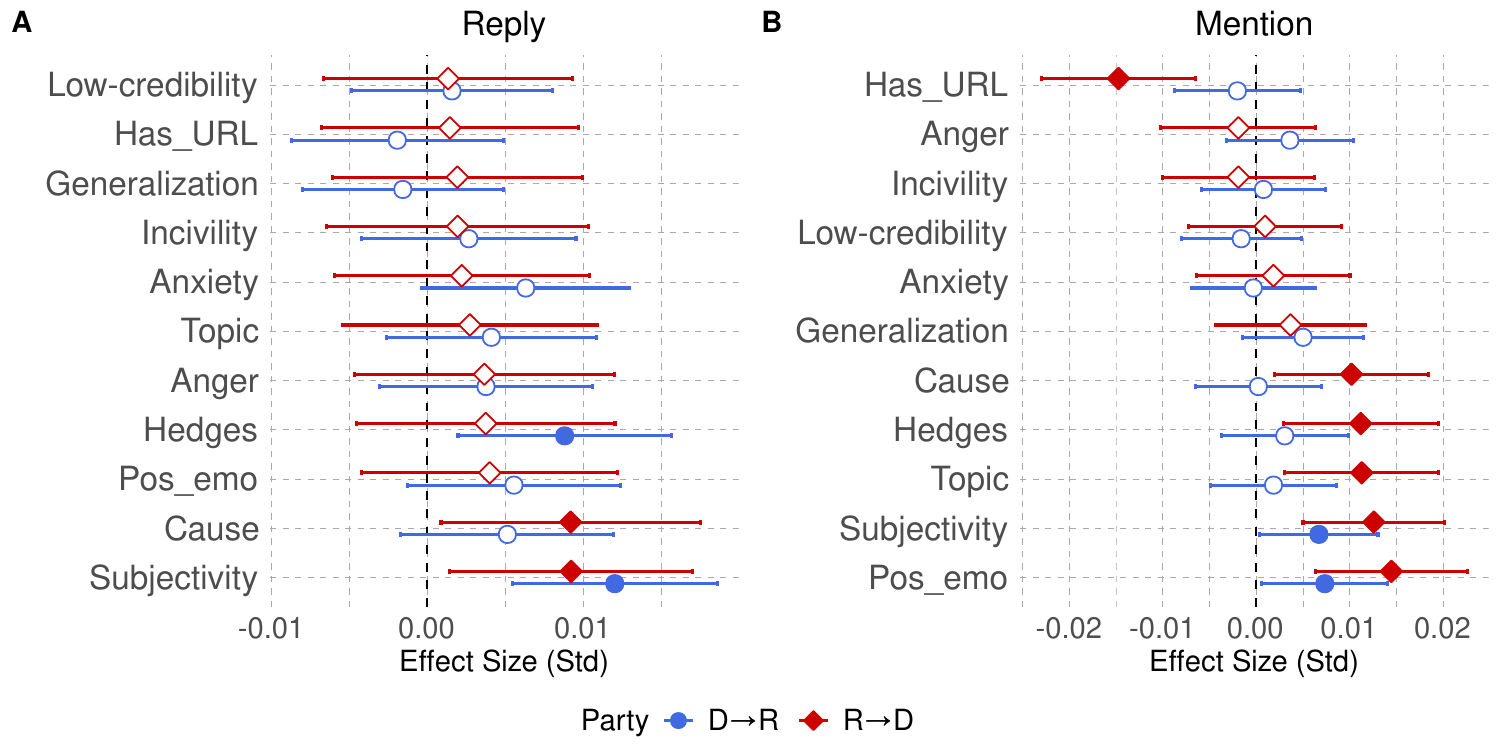}
\vspace{-0.5em}
\caption{\a{{\bf \b{Association between} \b{engagement} \b{and} future cross-talk style.} Prior CPI \b{engagement} \b{is associated with} differences in the rhetorical and stylistic choices employed by legislators in their subsequent cross-cutting interactions. Both Republicans and Democrats \b{exhibit shifts toward} more constructive CPI styles, as indicated by greater use of subjective, causal, and positive emotional language.}
} \label{fig:rq4}
\vspace{-1.5em}
\end{figure*}

\a{Figure~\ref{fig:rq4} presents how \b{engagement} gained from cross-party interactions \b{is associated with} the posting styles elites adopt in subsequent CPIs. We find systematic evidence that legislators \b{adjust} their linguistic and emotional strategies depending on whether CPIs previously attracted audience attention, \b{consistent with the interpretation that} \b{engagement} \b{is correlated with} stylistic differences in future cross-partisan communication.}


\a{\paragraph{Replies.}
In replies, parties \b{exhibit shifts toward styles characterized by greater subjectivity and hedging} following CPIs that receive higher \b{engagement}. For Democrats, \b{engagement is associated with} increased use of hedges ($\beta_L = 0.009, p < 0.05$) and subjective language ($\beta_L = 0.012, p < 0.001$), \b{which are linguistic markers commonly used to soften claims, express personal viewpoints, and signal epistemic uncertainty}. Republicans show a similar pattern, with greater use of subjective language ($\beta_L = 0.009, p < 0.05$) alongside more frequent use of causality terms ($\beta_L = 0.009, p < 0.05$). \b{Taken together, these patterns indicate that higher engagement is associated with replies that rely more on subjective framing and explanatory language, rather than categorical or declarative statements.}}

\a{\paragraph{Mentions.}
Mentions \b{are associated with} broader stylistic differences. Democrats are more likely to adopt subjective language ($\beta_L = 0.007, p < 0.05$) and express positive emotion ($\beta_L = 0.007, p < 0.05$) when prior mentions gain \b{engagement}, \b{corresponding to} more affectively warm and opinion-driven dialogue. Republicans \b{show stronger associations}: higher \b{engagement is associated with} reduced use of URLs ($\beta_L = -0.015, p < 0.001$), alongside increased use of hedges ($\beta_L = 0.011, p < 0.01$), subjective language ($\beta_L = 0.013, p < 0.01$), positive emotion ($\beta_L = 0.014, p < 0.001$), causality terms ($\beta_L = 0.01, p < 0.05$), and topical references ($\beta_L = 0.011, p < 0.01$). \b{These associations suggest that highly engaged mentions tend to co-occur with cross-party communication that is more conversational, affective, and narrative-driven, while relying less on external links.}}

\noindent \a{\paragraph{Main Takeaways.}
Across interaction types, we observe a consistent \b{association} between \b{engagement} and future posting styles in CPIs. Democrats \b{tend to exhibit} more subjective and emotionally positive cross-party communication, particularly in mentions and replies, while Republicans \b{display broader stylistic differences} that include causal reasoning and issue-centered framing. Importantly, in deliberative forms of CPI—replies and mentions—higher \b{engagement} \b{is systematically associated with} styles linked to more meaningful discourse: hedging and causal explanations that provide reasoning, subjective framing that signals personal engagement, and positive-emotional language.}

\a{We do not find evidence that \b{engagement is associated with increased} use of negative emotions such as anger, anxiety, or toxicity in future CPIs---despite these styles garnering high \b{engagement} for certain CPIs (RQ1b). This absence is noteworthy: while prior work raises concerns that engagement metrics may incentivize divisive or hostile rhetoric, our results indicate that, in the context of CPIs, \b{higher engagement tends to co-occur with} more constructive and deliberative communication strategies. \b{These patterns suggest that audience engagement is correlated with cross-party exchanges that are more deliberative, substantive.}}

\subsection{\x{Robustness Checks}}
\x{We perform several checks to assess whether our results are robust to alternative specifications and design choices. For RQ1, we try different matching strategies, including propensity-score and nearest-neighbor matching with varying covariate sets, and find consistent results across specifications. For RQ2 and RQ3, we reestimate models using monthly and biweekly instead of weekly aggregates, and the association between \b{engagement} and both the frequency and style of subsequent CPIs remains unchanged. We also test alternative mixed-effects model specifications, varying autoregressive terms, user-level random effects, and time fixed effects. \b{These checks yield results that closely mirror the main analyses, with no changes in the direction of effects and only minor variation in magnitude (within 2-5\% of reported results).} These checks suggest that our findings are not artifacts of modeling or sampling choices but reflect robust patterns in how \b{engagement} dynamics shape elite CPI.}

\vspace{-.5em}

\section{Discussion}

\a{Our work reframes cross-partisan interactions as part of a dynamic feedback system in which elites and audiences co-produce the trajectory of bipartisan communication. Classic elite cue theory emphasizes how politicians shape mass opinion; work on online influencers shows how highly visible figures adapt to audience signals~\cite{tufekci2013not}. Our findings bring these strands together: on social media, elites are not only cue-givers—they are also cue-takers, updating both whether and how they engage across party lines in response to audience \b{engagement}. \b{This perspective shifts the motivating question of whether CPIs depolarize from a static outcome-based framing to a dynamic process-based one: under what audience conditions do elites sustain, adapt, or abandon cross-partisan engagement over time?}}

\a{First, we show that \b{engagement} effects are asymmetric across parties and interaction types. Republican audiences tend to penalize direct forms of cross-party dialogue (mentions and replies), whereas Democratic audiences reward them with modest \b{engagement} gains. This asymmetry is important because it sets the stage for feedback: elites quickly learn which kinds of interactions work with their base. In doing so, CPIs become subject to the same attention dynamics as influencer communication more broadly, where audience signals guide strategic adjustments in style and frequency. \b{From a depolarization perspective, this suggests that CPIs are not uniformly encouraged or discouraged by audiences, but are instead selectively reinforced in ways that may shape the feasibility of sustained cross-party dialogue.}}

\a{Second, legislators who receive engagement on CPIs are more likely to continue such exchanges in subsequent weeks, and they increasingly frame these interactions with causal reasoning, subjective expression, topical engagement, and even positive emotion. Notably, we do not observe reinforcement of negative emotions such as anger, anxiety, or toxicity. This runs counter to concerns that online attention invariably amplifies divisive rhetoric. \b{While our study does not directly measure changes in audience attitudes or affective polarization, these patterns suggest that engagement-driven feedback does not necessarily push elites toward more hostile or inflammatory cross-party communication. Instead, when CPIs are sustained, the feedback loop of \b{engagement} appears to favor styles associated with explanation, perspective-taking, and issue-focused discourse—key preconditions often theorized to support deliberative and depolarizing processes.}}

\a{\b{These findings nuance the motivating question of whether cross-partisan interactions depolarize. Rather than treating CPIs as inherently polarizing or depolarizing acts, our results suggest that their longer-term implications depend on how audiences respond to them and how elites adapt in turn.}}\a{This highlights both a challenge and an opportunity for platform governance. If left unchecked, engagement metrics can reinforce antagonism within partisan silos. But when applied to cross-partisan contexts, they may encourage elites into more constructive forms of dialogue. Designing healthier digital public spheres will therefore require recognizing this dual potential of feedback loops—amplifying destructive behaviors in some cases~\cite{caplan2016controls}, but also incentivizing elites to experiment with and sustain more issue-focused, relational, and positively framed CPIs. \b{Platform interventions that differentially surface or contextualize cross-partisan engagement may play a critical role in shaping whether CPIs contribute to polarization dynamics or help mitigate them over time.}}


\subsubsection{Limitations and Future Work.} 
Our analysis provides new evidence of feedback loops in elite cross-party communication, but certain limitations remain. We focus on Twitter during 2020–21, a period shaped by specific platform affordances and events; patterns of \b{engagement} and reinforcement may differ elsewhere. While we measure \b{engagement} relative to legislators’ baselines, the lack of granular audience identifiers limits our ability to separate co-partisan from cross-partisan drivers. \b{Importantly, our analyses are observational and describe associations, and we do not claim any causal effects. As with most large-scale observational work, our findings should be interpreted as descriptive associations, with multiple robustness checks supporting the stability of these patterns across alternative specifications.} 

Future work should extend this analysis across platforms, election cycles, and political systems to test the generalizability of feedback effects. Experimental and mixed-method approaches could further clarify how audiences interpret and reward rhetorical strategies, deepening understanding of when engagement fosters bridging vs. division.

\section*{Acknowledgements}
The authors would like to acknowledge support from AFOSR, ONR, Minerva, NSF \#2318461, and Pitt Cyber Institute's PCAG awards. The research was partly supported by Pitt’s CRCD resources. Any opinions, findings, and conclusions or recommendations expressed in this material do not necessarily reflect the views of the funding sources.

\bibliography{main}


\clearpage
\section*{\ahb{Appendix}}

\b{Note: For all manual labeling tasks reported in this paper, annotations were conducted independently by two annotators, and final labels were determined through discussion to resolve any disagreements.}

\paragraph{Ethical Considerations.} 
We collected data from Twitter using Twitter's Official API v2.0 before rate limitations were imposed. The data are posted by public figures and available for viewing without any restrictions. The study is performed at an aggregate level and we do not analyze or report any results at the individual level. Examples of actual tweets are only shared in Table~\ref{tab:rq2_example} where we removed all personally identifiable information. \ahb{Moreover, the results and implications presented in this work should not be interpreted by political elites as a means to boost their engagements on social media.} 

\ab{\paragraph{Engagement metric for retweets.} Engagements on retweets are influenced by the engagement of the original post and its author. The engagement metric for retweets are limited to retweet counts, which inherently reflect the popularity of the original post. This limitation arises because the Twitter API attributes retweet engagement metrics (i.e., retweet counts) to the original post rather than to the retweeted post itself. Given these constraints, it is not feasible to fully disentangle the independent engagement dynamics of retweets versus the original posts. However, as legislators often have large follower bases, it is likely that their retweets amplify the \b{engagement} of the original posts, making the study of cross-party retweets an important part of our study.} 

\ab{The use of the overperforming score metric (Section~\ref{sec:vis}) mitigates some of the concerns regarding the inability to separate retweet and original post engagements by ensuring that engagements (even across retweets) are analyzed relative to a legislator’s baseline engagement levels. We analyze the correlation between the mean overperforming scores of a legislator's original posts and their retweeted posts, finding a significant positive Pearson correlation (0.38, $p<0.05$). While modest, this correlation demonstrates that engagement dynamics for retweeted posts align with those of a legislator’s original posts, relative to their baseline engagement levels. This alignment suggests that retweeted posts are not outliers or fundamentally different in terms of engagement patterns. Therefore, the overperforming score metric remains a valid tool for capturing engagement dynamics across post types. Although the metric in terms of retweets does not fully isolate the direct influence of the original post's engagement, the significant result asserts minimal biases from external factors, providing a reliable basis for analysis.
}



\paragraph{Cross-cutting interactions.} We manually label 300 followers\footnote{\a{The median number of followers per legislator is 489 (with maximum following of 24,598). We report median since the distribution of followers is skewed.}} for their political leaning (i.e., Democrat vs. Republican)  using stratified sampling \b{that intentionally oversamples users with more extreme attention-leaning scores (i.e., above 0.5 and below -0.5). Our goal is not to infer partisan leaning for all users, but to calibrate reliable thresholds for identifying users with clear Republican or Democratic leanings. Users near the center of the distribution frequently follow legislators from both parties and often lack a consistent partisan stance in their own content, making them inherently ambiguous and unsuitable for threshold calibration.} \ab{During the manual labeling process, we analyze (at least five) recent posts per account to determine the political leaning of the users. An account is labeled as Republican or Democrat if the majority of its posts express a clear stance supporting or opposing a specific party, policy, or ideology. If the posts lack a consistent political stance or focus on non-political topics, the account is classified as non-partisan or apolitical. This likely ensures that the classifications are grounded in the user's behavior, reducing the likelihood of misclassifications due to ambiguous following patterns.} 

\begin{figure}[ht]
\setlength{\tabcolsep}{0pt}
\begin{tabular}{cc}

\includegraphics[width=0.5\linewidth]{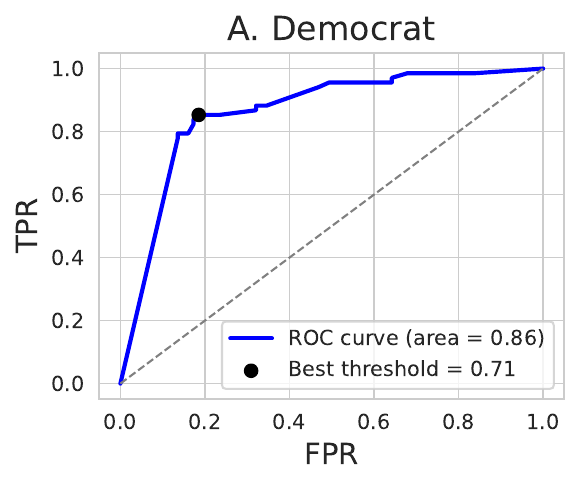}
&
\includegraphics[width=0.5\linewidth]{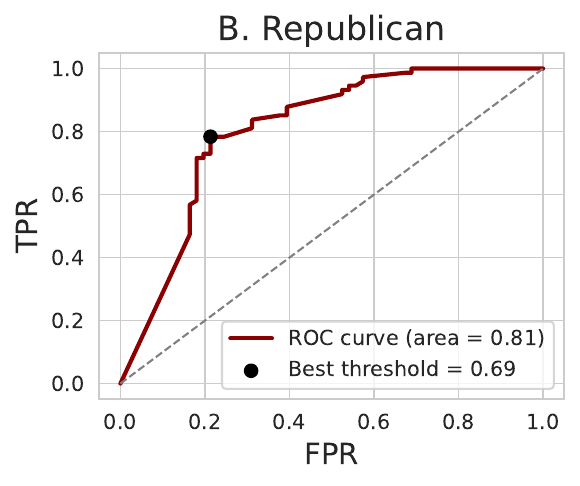}

\end{tabular}
\caption{ROC curves for (A) Democrats and (B) Republicans showing optimal thresholds to determine user leanings.} 
\label{fig:roc}
\end{figure}

Two annotators labeled the samples with a Cohen Kappa score of 0.78 (substantial agreement). The final labels were decided by mutual discussion. Based on the ROC curves (Figure~\ref{fig:roc}), we set the threshold \ab{(by maximizing TPR and minimizing FPR simultaneously)} for identifying a follower as Democrat (AUC = 0.86) at 0.71 (i.e., scores between 0.71 and 1) and as Republican (AUC = 0.81) at -0.69. \ab{For Democrat leaning accounts, the precision, recall and F1 scores are 0.79, 0.85 and 0.82 respectively at the chosen cutoff. For Republican leaning accounts, the precision, recall and F1 scores are 0.81, 0.78 and 0.80 respectively.}

\begin{figure}[ht]
\centering

\includegraphics[width=0.9\linewidth]{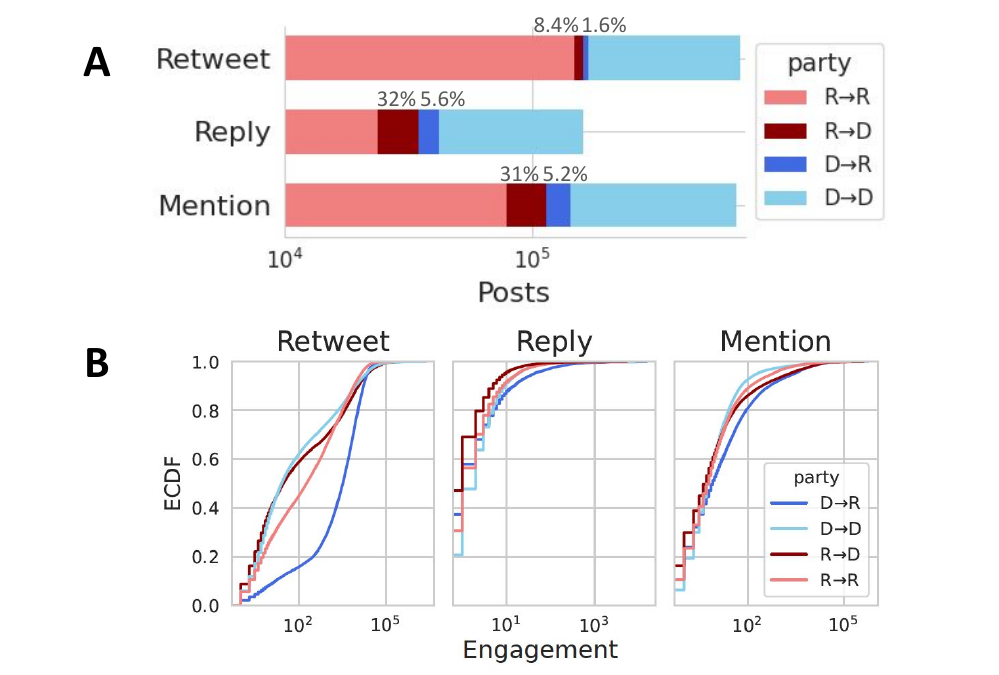}\\

\vspace{-0.5em}
\caption{{\bf Distribution of (A) cross-party posts and their (B) engagement.} The rate of cross-talk is higher for Republicans compared to Democrats across all types of interactions. However, typical engagements on cross-party posts tend to be lower than intra-party ones for Republicans.}
\label{fig:data}
\vspace{-1.5em}
\end{figure}

\ab{\paragraph{Distribution of cross-party posts and their engagement.} Figure~\ref{fig:data} shows the distribution of cross-cutting and intra-partisan posts and their engagement.}

\paragraph{Central (C) vs. local (L).} We identify the location of the legislator's followers using the user location field returned by the Twitter API. This field is user-generated containing noisy data. We use regular expressions to identify user's state if they are from the US else their country. Around 46\% followers are mapped to the states. The locality of a legislator's followers is measured as the proportion (calculated out of the followers having a location) of followers belonging to the same state as the legislator. The distribution of local and central legislators are similar for Republicans and Democrats, suggesting that no (or minimal) biases are introduced for party using our categorization method.

\ab{\paragraph{\b{Measuring Engagement}.} We borrow the \b{engagement} metric and the suggested parameters $b_0=10$ and $m=14$, from \citet{biswas2025political}, who demonstrated these parameters to be effective in capturing meaningful engagement variations. 
The $b_0$ is based on the daily mean engagement received by legislators on Twitter. It serves as a threshold to filter out noisy engagement posts by ensuring that only posts with substantial engagement levels are considered, thereby reducing the impact of low-engagement anomalies on the \b{engagement} metric.
The ideal window $m$ is estimated based on legislators' daily posting rates on these platforms. Choosing $m=14$, ensures a reliable estimate of expected engagement $b_u$ by incorporating a sufficient number of posts per legislator while accounting for temporal variations in posting behavior and audience activity.}

\ab{These design choices are likely to make the \b{engagement} metric robust and meaningful across varying levels of activity among legislators. While the confidence in the median engagement $b_u$ may be lower for legislators with fewer posts, this is likely mitigated by the choice of $m=14$, which is derived from the typical posting rates of legislators. This ensures that, on average, we have a sufficient sample size per legislator for a reliable calculation of $b_u$. Additionally, our analysis does not rely on the cumulative \b{engagement} of all posts but instead focuses on the median engagement for each legislator during the preceding window. This approach inherently controls for variations in posting frequency, as it standardizes engagement levels across legislators regardless of their activity rates, reducing potential biases introduced by differences in the volume of posts.
}

\b{\paragraph{Precision of Linguistic Markers.} 
We manually label 40 random samples per linguistic marker (i.e., 280 posts overall) to evaluate the quality of our lexicon-based features. Annotators showed high agreement (Cohen's $\kappa$ = 0.82), with disagreements resolved through discussion to produce final reference labels. Table~\ref{tab:prec_lm} reports the precision of the lexicon-based indicators with respect to these human-validated labels, both overall and by party. The overall precision ranges between 0.625-0.925, indicating that when a linguistic marker is identified by the lexicon-based method, it typically corresponds to the intended construct as judged by human annotators. We focus on precision rather than recall because our goal is to use conservative, high-confidence indicators of linguistic traits at scale, rather than to exhaustively capture all possible instances of a given style. The precision of certain markers, such as \texttt{cause} and \texttt{pos\_emo}, differs across parties, which may reflect differences in communication styles.
}

\b{\textit{LLM-based validation of posting style annotations.}
To further assess the reliability of our lexicon-based posting style indicators, we conducted an additional validation using a large language model (LLM). For these 280 posts, we prompted LLaMA-3-70B (Instruct) to judge whether each post exhibited specific linguistic characteristics (e.g., subjectivity, causality, hedging, incivility) using the same definitions as our lexicon-based measures. We then compared the LLM’s judgments with the binary labels produced by the lexicon-based approach. We observe good agreement between the two methods across most posting styles, as reported in Table~\ref{tab:prec_lm}. Disagreements primarily arise in cases where pragmatic interpretation is required. For instance, the LLM sometimes infers sarcasm, implied blame, or emotional intent in posts that contain no explicit linguistic markers, leading to false positives. In contrast, the lexicon-based approach is more conservative, labeling a post only when explicit lexical cues are present. Given these trade-offs, we retain lexicon-based indicators as our primary measures for scalability and interpretability.}

\begin{table}
    \centering
    \scriptsize
    \setlength{\tabcolsep}{1.5pt}
    \begin{tabular}{lccccccc}
    \toprule
         & Anger & Anxiety & Cause & Generalization & Hedges & Pos\_emo & Subjectivity\\
         \midrule
        Overall (Manual) & 0.625 & 0.775 & 0.725 & 0.925 & 0.675 & 0.775 & 0.825 \\
        Dem (Manual) & 0.643 & 0.806 & 0.667 & 0.923 & 0.620 & 0.710 & 0.786\\
        Rep (Manual) & 0.583 & 0.667 & 0.900 & 0.928 & 0.818 & 1.000 & 0.917\\
        \b{LLaMA-3-70B} & \b{0.731} & \b{0.782} & \b{0.820} & \b{0.846} & \b{0.680} & \b{0.825} & \b{0.715} \\
        \bottomrule
    \end{tabular}
    \caption{Precision of Linguistic Markers}
    \label{tab:prec_lm}
\end{table}

\paragraph{Topics.} \ab{Our topic selection approach is guided by the study's focus on understanding how salient topics in cross-party discussions affect the \b{engagement} of legislators, rather than on an exhaustive exploration of all possible topics in cross-party discussions. The topics selected (BLM, COVID-19, rights, immigration, gun control, climate, abortion, and the Capitol riots) represent highly salient and contentious issues in contemporary U.S. politics. While not exhaustive, this selection ensures coverage of areas that are critical for analyzing US political discussions. By focusing on these topics, we aim to capture the dynamics of cross-party interactions around some of the most significant issues of public and political interest.} 

\ab{We use a keyword-based approach for initial topic detection because it offers interpretability and ensures high precision. This method is particularly effective given the brevity of tweets, which often limits the performance of unsupervised techniques like BERTopic. While we initially experiment with BERTopic to identify topics in an unsupervised manner, the results are noisy (and especially ineffective due to our focus on a specific set of topics), likely due to the short and informal nature of tweets. Consequently, we adopt a more structured and precise methodology to identify topical tweets.}

\ab{An initial set of keywords (shown in Table~\ref{tab:topic}) specific to each topic is used to identify topical tweets. While this set may not be comprehensive, it is designed to maximize precision over recall, thereby reducing false positives. This ensures that the posts identified for a topic are highly relevant, even if some relevant posts are excluded. Tweets identified for each topic are clustered based on their RoBERTa embeddings (using Nearest Neighbor) to capture semantic similarity. This step gives clusters of posts under each topic which is used to identify relevant clusters for each topic. We manually label 25 posts per cluster to identify and exclude noisy clusters under each topic. The clusters containing posts irrelevant to the topic are filtered out. Finally, we assign all posts belonging to relevant clusters to the seed topic. An additional 50 posts per topic (after filtering out noisy clusters) are manually labeled to ensure the relevance of classified posts. The precision of our topic labeling ranges between 84-100\%, demonstrating the reliability of the approach. Based on our methodology, we are able to assign posts to multiple topics despite the short text, enabling nuanced analysis of complex discussions. Table~\ref{tab:topic} shows the distribution of topics for cross-cutting and co-partisan posts\footnote{Previously, the topic `race' was also included  but the clusters for this topic were noisy, hence we removed it for further analysis}.
}


\ab{\paragraph{Low-credibility.}
Among the low-credibility URLs\footnote{\ab{Around 11.1\% of posts include URLs. We are unable to determine how many of these URLs are news-related, as recovering full URLs for all posts is non-trivial. This limitation arises because URLs in Twitter posts are often truncated, requiring additional data collection or third-party services to expand and categorize them reliably.}}, 38.1\% are news-related. Using URLs to detect low-credibility content is a common practice in the literature~\cite{lasser2022social}, particularly for short texts like tweets, as it provides a straightforward method to evaluate the reliability of shared information. While other methods, such as detecting fact-checked claims using LLMs, are promising, they present challenges in our context. Our dataset spans an older time period, during which fact-checking datasets may be sparse. Furthermore, LLM-based approaches risk hallucination and lack robust interpretability, making validation without proper ground truth data more difficult. This underscores the practicality of our current approach, although future work could explore integrating newer methodologies as tools and datasets improve.
}

\ab{It is important to note that our work is not focused on quantifying the prevalence of low-credibility content overall. Instead, it explores the role of low-credibility content as a specific posting style in the \b{engagement} dynamics of cross-cutting interactions. Thus, we are only interested in how the use of such content may influence interactions between opposing partisan groups, rather than the amount or nature of low-credibility content legislators share.
}

\begin{figure*}[ht]
\centering
\includegraphics[width=\linewidth]{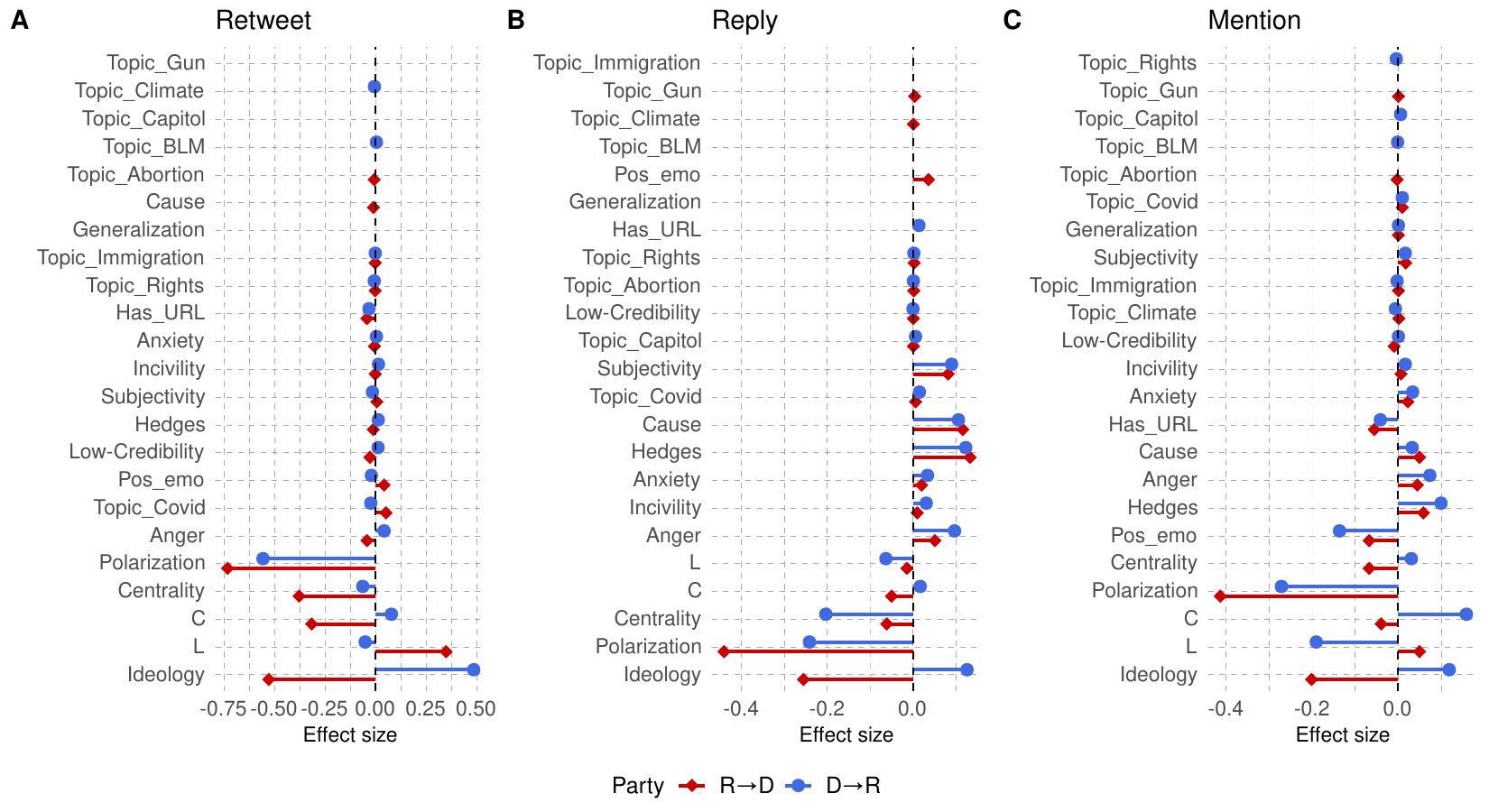}
\vspace{-0.5em}
\caption{
\yrv{{\bf Differences in Posting Styles and Individual Characteristics.} The posting styles and individual attributes of legislators vary between cross-party and intra-party posts, for both political parties and across both interaction types. The figure uses lollipop charts to show the differences between cross-party and intra-party (baseline) posts, measured by effect sizes through the Mann-Whitney U test. Only statistically significant effects ($p < 0.05$) are displayed. \ab{The covariates are sorted based on differences in effect sizes between Republicans and Democrats (low to high). Effect size sign: (+) More frequent in cross-party group; (-) More frequent in intra-party group.}}
} \label{fig:mwu}
\end{figure*}

\begin{figure*}[ht]
\centering
\includegraphics[width=0.9\linewidth]{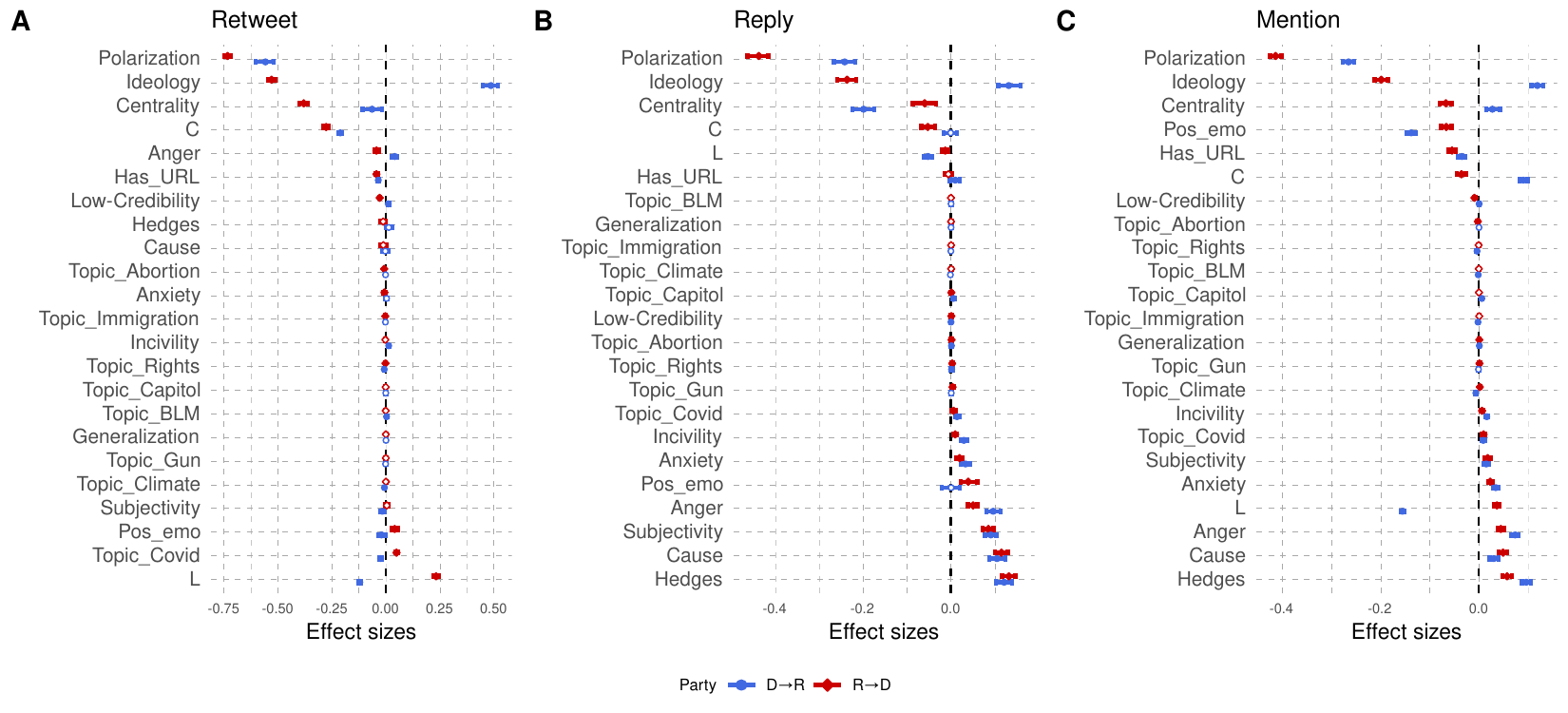}
\vspace{-.5em}
\caption{Confidence intervals of Mann-Whitney test (Figure~\ref{fig:mwu}). The covariates are sorted based on effect sizes for Republicans.} \label{fig:rq0_ci}
\vspace{-1.5em}
\end{figure*}

\paragraph{Mann-Whitney U test.} Figure~\ref{fig:mwu}) shows the Mann-Whitney effect sizes for differences in cross-cutting vs. intra-party posts (Figure~\ref{fig:rq0_ci} shows the confidence intervals). 

\begin{figure}[ht]
\centering
\setlength{\tabcolsep}{3pt}
\begin{tabular}{c}

(a) D$\rightarrow$R \\ \includegraphics[width=0.95\linewidth]{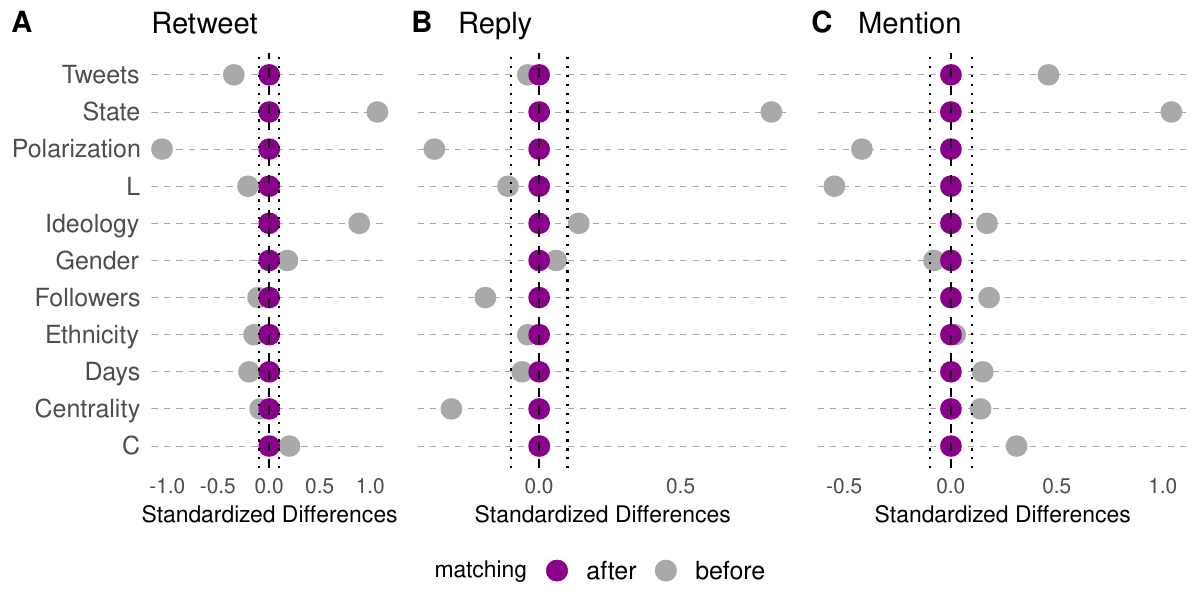}
\\
(b) R$\rightarrow$D \\ \includegraphics[width=0.95\linewidth]{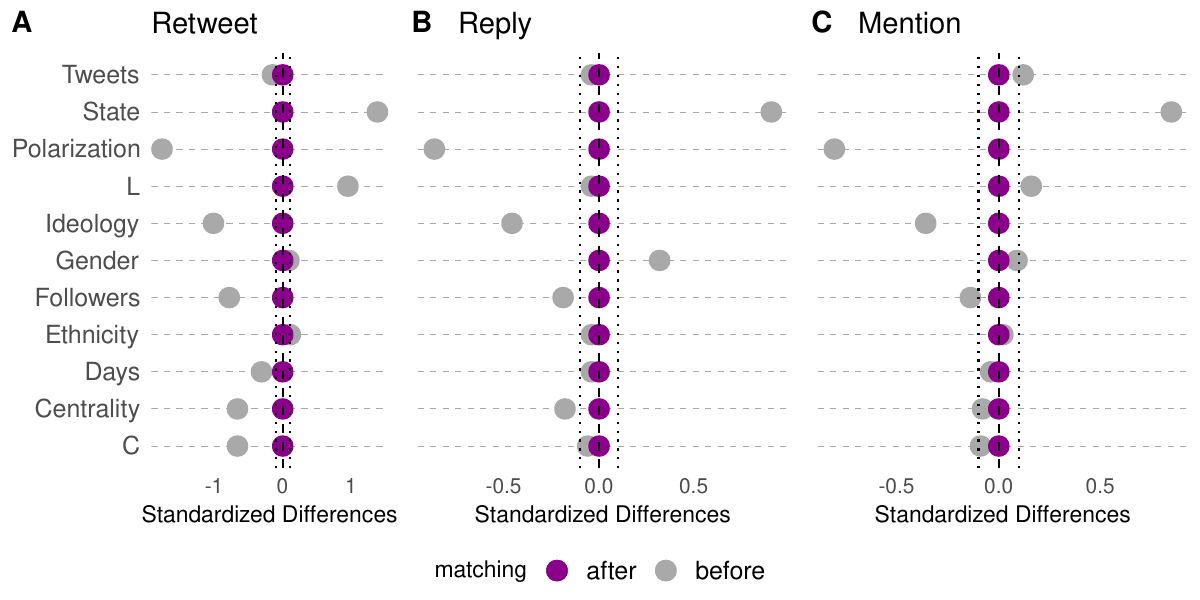}\\

\end{tabular}
\caption{{\bf Covariate balance for matching.} The standardized differences for each of the covariates before and after matching for cross-cutting posts by (a) Democrats and (b) Republicans. All the covariates are balanced (i.e., score between -0.1 to 0.1) after matching in each case.}
\label{Fig:2}
\end{figure}

\begin{figure}[ht]
\centering
\setlength{\tabcolsep}{3pt}
\begin{tabular}{c}

(a) D$\rightarrow$R \\ \includegraphics[width=0.95\linewidth]{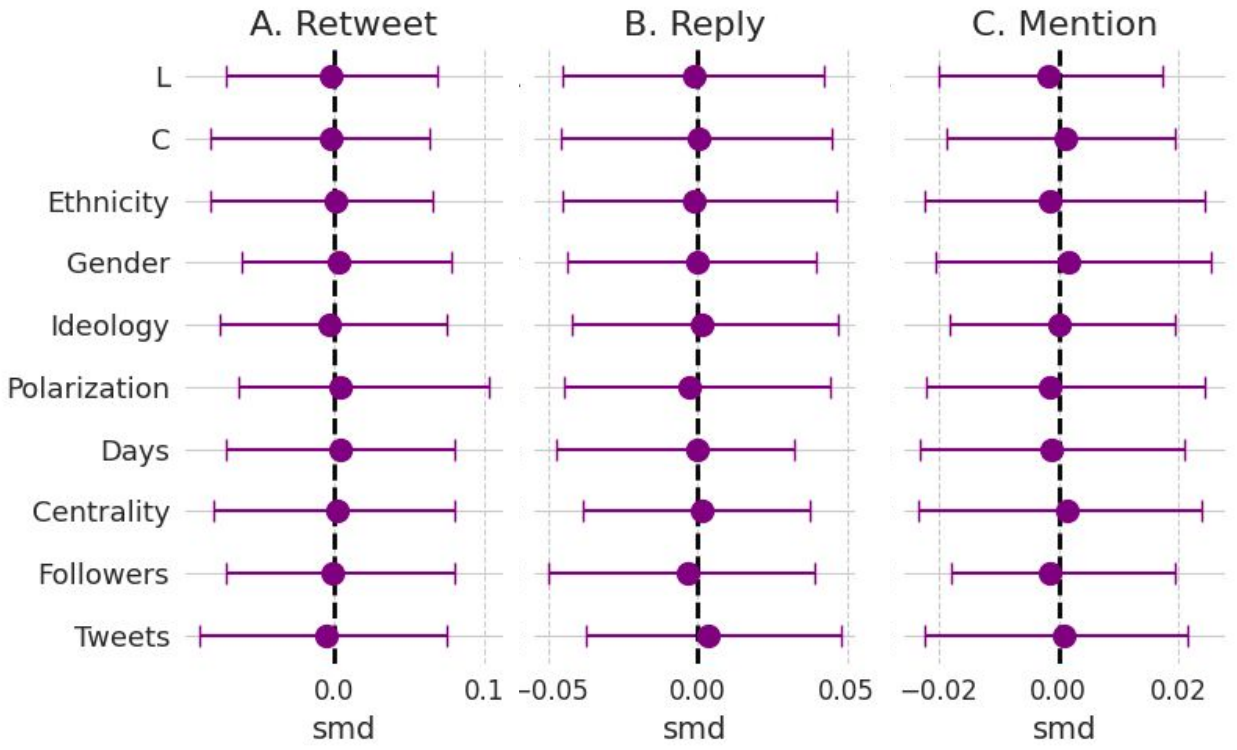}
\\
(b) R$\rightarrow$D \\ \includegraphics[width=0.95\linewidth]{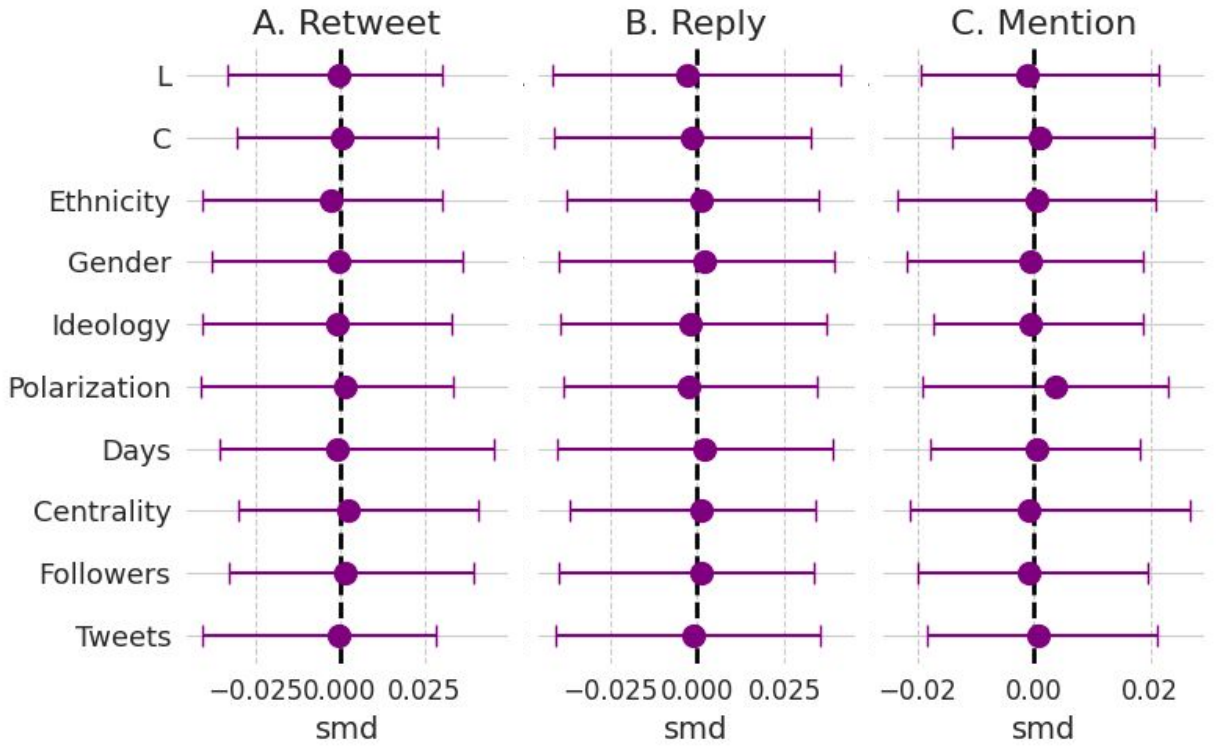}\\

\end{tabular}
\caption{\ab{{\bf Distribution of Standardized Differences \yrvv{for Covariates}.} The mean and 95\% confidence intervals of standardized mean differences (smd) for covariates after matching for cross-cutting posts by (a) Democrats and (b) Republicans. The feature `State' is not shown in the analysis due to its sparsity, which leads to multicollinearity issues in the calculation of standardized differences.}}
\label{Fig:2.1}
\end{figure}
\paragraph{Matching.} \b{We employ 1:1 matching using K-nearest neighbor algorithm (with Euclidean distance) for RQ1a and 1b. The distance cutoff is chosen such that the covariate overlap between treated and control groups are maximized. For cross-cutting posts by Republicans we find matches for 73.2\% retweet, 74.7\% reply, and 75.3\% mentions. For Democrats, we find matches for  78.2\% retweet, 76.2\% reply, and 78.8\% mentions. All the covariates are balanced after matching as shown in Fig.~\ref{Fig:2} \ab{(bootstrapped mean and 95\% confidence intervals for standardized differences after matching are shown in Fig.~\ref{Fig:2.1}).}}

\ab{\paragraph{Data Transformation for RQs.} The variables were transformed to be close to normal distributions to satisfy the model assumptions. For RQ1a and 1b, the transformations are: $y$ (log), audience partisanship (arcsinh), centrality (center+scale), \#posts (log), \#followers (Yeo-Johnson), and ideology (arcsinh). For RQ2 and 3, the transformations are: $\rho$ (log), $\delta$ (arcsinh), ideology (sqrt), \#followers (arcsinh), \#posts (Yeo-Johnson), centrality (Yeo-Johnson), and audience partisanship (Yeo-Johnson).}





\begin{figure}[ht]
\vspace{-1.5em}
\centering
\setlength{\tabcolsep}{3pt}
\begin{tabular}{c}

(a) \\ \includegraphics[width=0.7\linewidth]{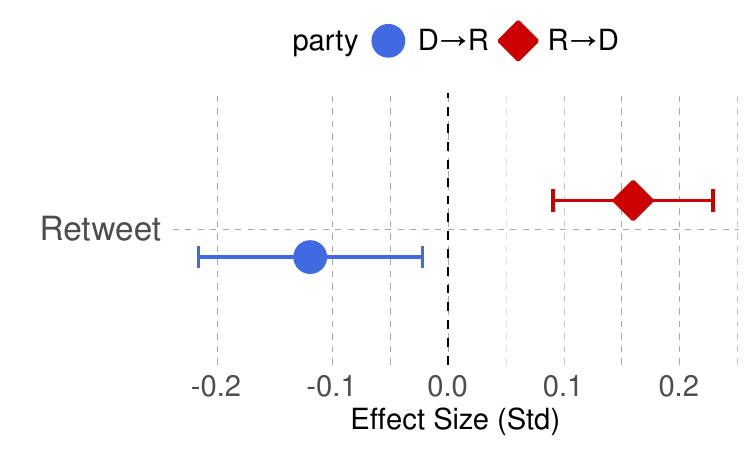}
\\
(b) \\ \includegraphics[width=0.75\linewidth]{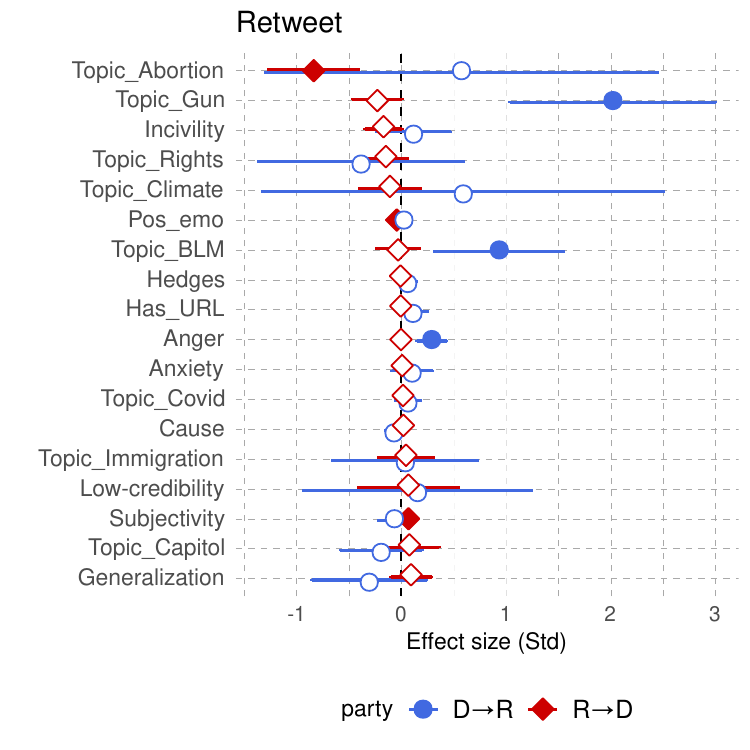}

\end{tabular}
\caption{Effect of (a) cross-talk (RQ1a) and (b) corresponding posting styles (RQ1b) on \b{engagement} for Retweet.}
\label{fig:rq_ret}
\end{figure}

\a{\paragraph{Effect of cross-talk for Retweet.} Figure \ref{fig:rq_ret} shows the effect of cross-talk on \b{engagement} (RQ1) for Retweet.}

\a{As shown in Figure~\ref{fig:rq_ret}(a), Republicans get higher \b{engagement} when they retweet Democrats ($\alpha_1$ = 0.16, $p<0.001$) i.e., a post tends to overperform by 16\%, on average, if a Republican legislator retweets the post by a Democrat. However, Democrats get lower \b{engagement} on retweeting Republicans ($\alpha_1$ = -0.12, $p<0.05$). This suggests that Republican audiences respond more positively to amplifying cross-cutting content, leading to increased \b{engagement} when Republicans engage with Democratic posts. On the other hand, Democratic audiences tend to engage less with cross-cutting content, leading to decreased \b{engagement} when Democrats retweet Republican posts. This asymmetry reflects the differences in how the two groups interact with content that presents opposing viewpoints.}

\a{Figure~\ref{fig:rq_ret}(b) shows that topics like abortion ($\alpha_L = -0.84$) and gun control ($\alpha_L = -0.23$) reduce the \b{engagement} of Republicans when they retweet Democrats. In contrast, Democrats gain higher \b{engagement} when retweeting Republicans on issues such as gun control ($\alpha_L = 2.02$) and BLM ($\alpha_L = 0.93$). One possible explanation is that Democrats' retweets tend to focus more on specific issues (e.g., sample tweets \circled{4} and \circled{6} in Appendix Table~\ref{tab:rq2_example})), while Republicans are more likely to retweet Democratic content sarcastically (e.g., \circled{1} and \circled{3} in Appendix Table~\ref{tab:rq2_example}). Notably, the presence of anger \ab{is associated with} higher \b{engagement} for Democrats ($\alpha_L$ = 0.29). These cross-party retweets tend to be stances expressed about certain issues (e.g., \circled{10} in Appendix Table~\ref{tab:rq2_example}), suggesting that such posts resonate with Democratic audiences.}

\begin{figure}[ht]
\centering
\setlength{\tabcolsep}{3pt}
\begin{tabular}{c}

(a) \\ \includegraphics[width=0.7\linewidth]{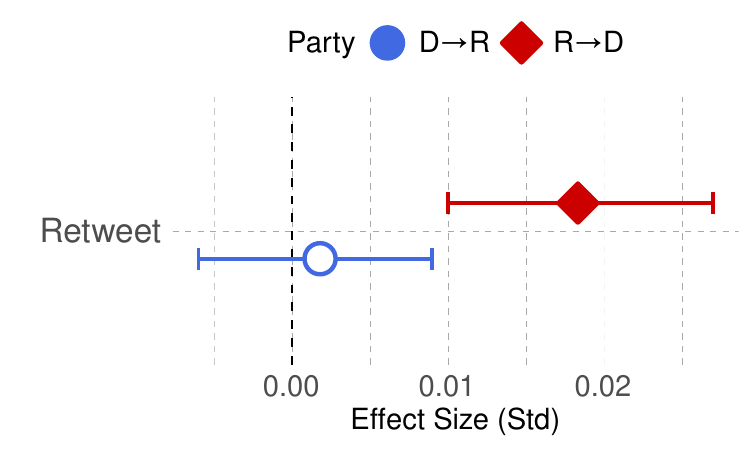}
\\
(b) \\ \includegraphics[width=0.75\linewidth]{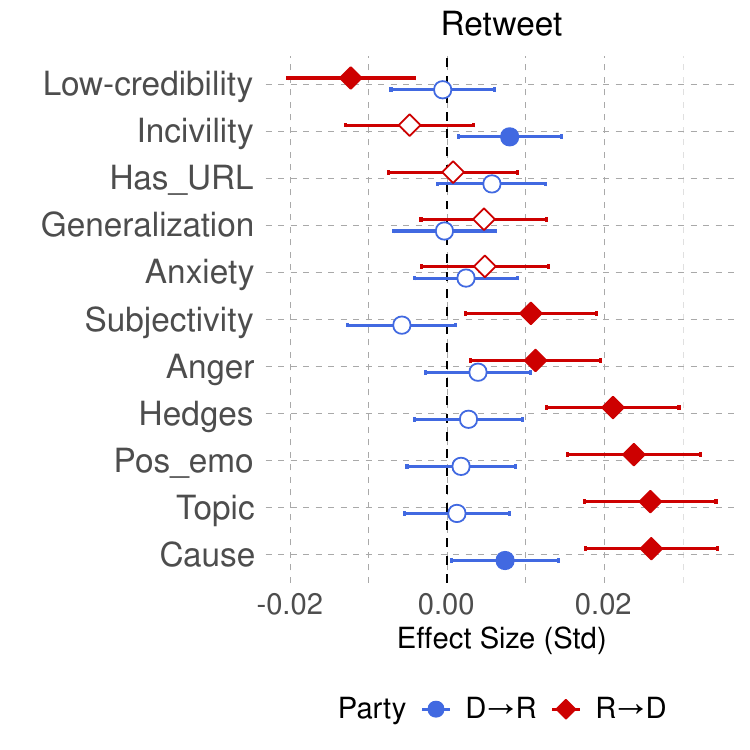}

\end{tabular}
\caption{Effect of \b{engagement} on future (a) cross-talk rate (RQ2) and (b) stylistic choices (RQ3) for Retweet.}
\label{fig:rq_ret_fl}
\end{figure}

\a{\paragraph{Feedback Loop Effect for Retweet.} Figure \ref{fig:rq_ret_fl} shows the effect of \b{engagement} on future cross-talk rate (RQ2) and stylistic choices (RQ3) for Retweet.}
\a{Republican legislators are more likely to retweet Democrats by 2.2\% ($\beta_1$ = 0.018, $p<0.001$) in the future when cross-party retweets gain higher \b{engagement} as shown in Figure~\ref{fig:rq_ret_fl}(a). However, no significant effects are seen for Democrat elites. This is aligned with the results of RQ1 which suggests that retweets are a preferred mode of cross-partisan interaction for Republican audiences, unlike Democrats. Therefore, the gain in \b{engagement} from cross-party retweets motivates Republican legislators to further engage in such interactions in the future.}

\a{Retweets show the strongest evidence of stylistic reinforcement among Republicans as shown in ~\ref{fig:rq_ret_fl}(b). For Democrats, visible CPIs lead to increased use of toxicity ($\beta_L = 0.0080, p < 0.05$) and causality terms ($\beta_L = 0.0074, p < 0.05$), suggesting that \b{engagement} pushes them toward sharper and more explanatory retweeting of opponents. Republicans exhibit a wide array of reinforced styles: \b{engagement} predicts higher use of hedges ($\beta_L = 0.0211, p < 0.001$), subjective language ($\beta_L = 0.0107, p < 0.05$), positive emotion ($\beta_L = 0.0238, p < 0.001$), anger ($\beta_L = 0.0113, p < 0.01$), causality terms ($\beta_L = 0.0260, p < 0.001$), and topics ($\beta_L = 0.0259, p < 0.001$). Importantly, \b{engagement} is negatively associated with misinformation content ($\beta_L = -0.0123, p < 0.01$), suggesting that audiences discourage Republican elites from retweeting misleading or unreliable material. Together, these results show that retweets act as a particularly strong site of stylistic feedback loops, with Republicans especially likely to double down on emotional and issue-focused styles when their retweets gain traction.}

\paragraph{RQ1b: $p$-value and examples.} Table~\ref{tab:rq2_p} shows the regression coefficients and corresponding $p$-values for RQ1b. Table~\ref{tab:rq2_example} shows examples of some of the posting styles associated with the \b{engagement} of cross-cutting posts.



\begin{table*}
    \small
    \centering
    \setlength{\tabcolsep}{3pt}
    \caption{Keywords used to identify topics and distribution of topics for cross-partisan and co-partisan posts. \b{Percentages denote the share of posts in each interaction category containing the given topic.}}
    \label{tab:topic}
    \begin{tabular}{llrrrr}
        \toprule
        & & \multicolumn{2}{c}{\textbf{cross-partisan}} & \multicolumn{2}{c}{\textbf{co-partisan}} \\
        \textbf{Topic} & \textbf{\ab{Initial Keyword(s)}} & \textbf{R$\rightarrow$D}  & \textbf{D$\rightarrow$R} & \textbf{R$\rightarrow$R}  & \textbf{D$\rightarrow$D}\\
        \cmidrule(r){1-1} \cmidrule(rl){2-2} \cmidrule(l){3-4} \cmidrule(l){5-6}
        BLM & george, floyd, blacklivesmatter 
        & 88 (\b{0.19\%}) & 123 (\b{0.38\%}) & 489 (\b{0.22\%}) & 3,825 (\b{0.44\%})\\
        
        COVID-19 & covid, covid-19, pandemic 
        & 2,457 (\b{5.37\%}) & 1,826 (\b{5.59\%}) & 6,885 (\b{3.14\%}) & 50,743 (\b{5.79\%})\\
        
        rights & right(s) 
        & 287 (\b{0.63\%}) & 228 (\b{0.70\%}) & 1,128 (\b{0.51\%}) & 10,595 (\b{1.21\%})\\
        
        immigration & immigration, immigrating, immigrant(s) 
        & 108 (\b{0.24\%}) & 78 (\b{0.24\%}) & 687 (\b{0.31\%}) & 3,950 (\b{0.45\%})\\
        
        gun control & gun(s), shooting 
        & 313 (\b{0.68\%}) & 267 (\b{0.82\%}) & 1,139 (\b{0.52\%}) & 7,636 (\b{0.87\%})\\
        
        climate & climate 
        & 141 (\b{0.31\%}) & 74 (\b{0.23\%}) & 303 (\b{0.14\%}) & 8,373 (\b{0.96\%})\\
        
        abortion & abortion 
        & 1,733 (\b{3.79\%}) & 90 (\b{0.28\%}) & 1,733 (\b{0.79\%}) & 2,825 (\b{0.32\%})\\
        
        Capitol riots & capitol, riot(s) 
        & 76 (\b{0.17\%}) & 277 (\b{0.85\%}) & 413 (\b{0.19\%}) & 2,187 (\b{0.25\%})\\
        \bottomrule
    \end{tabular}
    \vspace{-0.5em}
\end{table*}

\begin{table*}[ht]
\setlength{\tabcolsep}{4pt}
\renewcommand{\arraystretch}{1}
\centering
\caption{Regression coefficients for RQ1b}\label{tab:rq2_p}
\begin{tabular}{lcccccc}
\toprule
 & \multicolumn{2}{c}{\textbf{Retweet}} & \multicolumn{2}{c}{\textbf{Reply}} & \multicolumn{2}{c}{\textbf{Mention}} \\
\cmidrule(r){2-3} \cmidrule(lr){4-5} \cmidrule(l){6-7}
 & R$\rightarrow$D & D$\rightarrow$R & R$\rightarrow$D & D$\rightarrow$R & R$\rightarrow$D & D$\rightarrow$R \\
\midrule
Abortion        & \textbf{-0.836}*** & 0.574 & \textbf{-0.291}† & 0.010 & \textbf{-0.232}** & 0.155 \\
BLM             & -0.030 & \textbf{0.935}** & 0.013 & 0.393 & -0.129 & -0.198 \\
Capitol riots   & 0.077 & -0.192 & -0.136 & \textbf{-0.891}*** & 0.166 & \textbf{0.251}*** \\
Climate         & -0.109 & 0.591 & -0.084 & 0.035 & 0.061 & -0.092 \\
Covid           & 0.016 & 0.063 & 0.014 & 0.047 & \textbf{0.131}*** & \textbf{0.127}*** \\
Gun             & \textbf{-0.229}† & \textbf{2.020}*** & 0.067 & 0.302 & 0.106 & \textbf{0.272}*** \\
Immigration     & 0.044 & 0.038 & -0.111 & -0.131 & 0.175 & -0.038 \\
Rights          & -0.147 & -0.384 & 0.004 & 0.279 & -0.112 & \textbf{0.275}** \\
Has\_URL        & -0.005 & 0.111 & \textbf{-0.154}** & -0.049 & \textbf{0.068}** & \textbf{0.253}*** \\
Incivility      & \textbf{-0.170}† & 0.117 & \textbf{0.181}* & -0.048 & \textbf{-0.210}*** & -0.075 \\
Generalization  & 0.092 & -0.306 & 0.246 & -0.043 & \textbf{-0.288}* & \textbf{0.315}* \\
Hedges          & -0.007 & 0.061 & -0.005 & 0.038 & \textbf{-0.114}*** & 0.001 \\
Subjectivity    & \textbf{0.068}** & -0.066 & -0.009 & \textbf{0.127}* & \textbf{0.323}*** & \textbf{0.301}*** \\
Pos\_emo        & \textbf{-0.043}*** & 0.025 & -0.005 & 0.019 & -0.015 & \textbf{-0.042}** \\
Anxiety         & 0.007 & 0.103 & \textbf{0.144}* & 0.047 & 0.032 & \textbf{0.076}* \\
Anger           & -0.003 & \textbf{0.290}*** & \textbf{0.095}* & 0.023 & \textbf{0.121}*** & \textbf{0.052}* \\
Cause           & 0.018 & -0.070 & -0.019 & 0.029 & \textbf{-0.034}* & 0.021 \\
Low-Credibility & 0.067 & 0.154 & -0.347 & -0.691 & 0.027 & 0.299 \\
\bottomrule
\end{tabular}

\vspace{0.5em}
\begin{tablenotes}
\footnotesize
\item † $p < 0.10$, * $p < 0.05$, ** $p < 0.01$, *** $p < 0.001$
\end{tablenotes}
\end{table*}








\begin{figure*}[ht]
    \centering

    \begin{subfigure}
        \centering
        \includegraphics[width=0.3\linewidth]{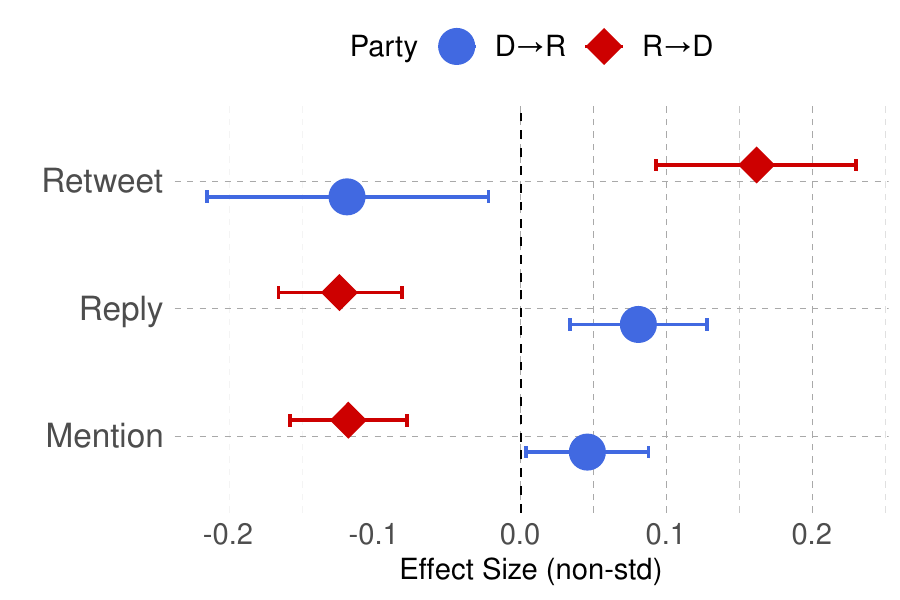}
        \caption{Non-standardized effect sizes of cross-cutting interactions on \b{engagement} (RQ1a).}
        \label{fig:rq1_ns}
    \end{subfigure}

    \vspace{1em} 

    \begin{subfigure}
        \centering
        \includegraphics[width=0.75\linewidth]{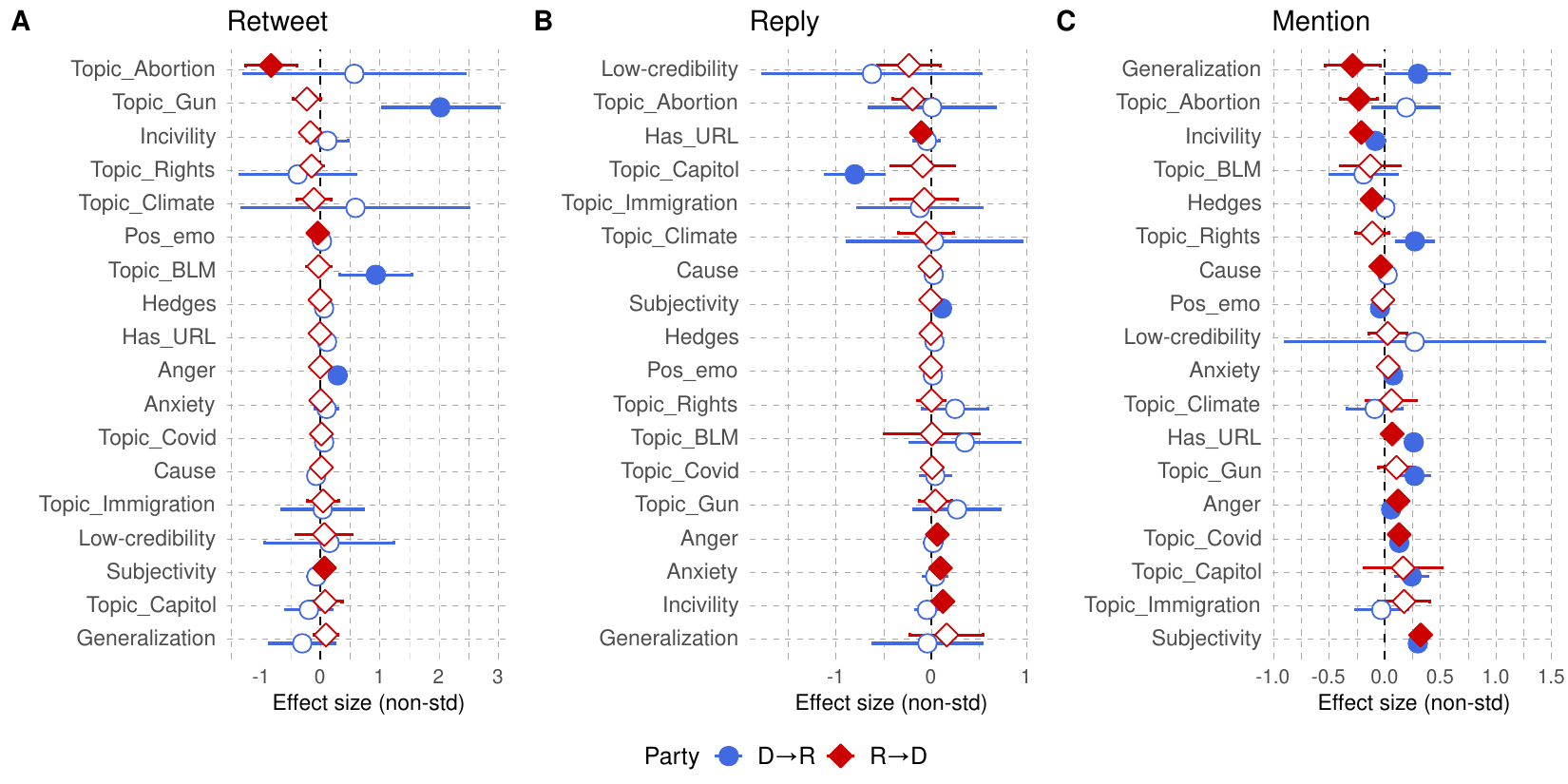}
        \caption{Non-standardized effect sizes of posting styles on \b{engagement} of cross-cutting interactions (RQ1b).}
        \label{fig:rq2_ns}
    \end{subfigure}

    \vspace{1em}

    \begin{subfigure}
        \centering
        \includegraphics[width=0.3\linewidth]{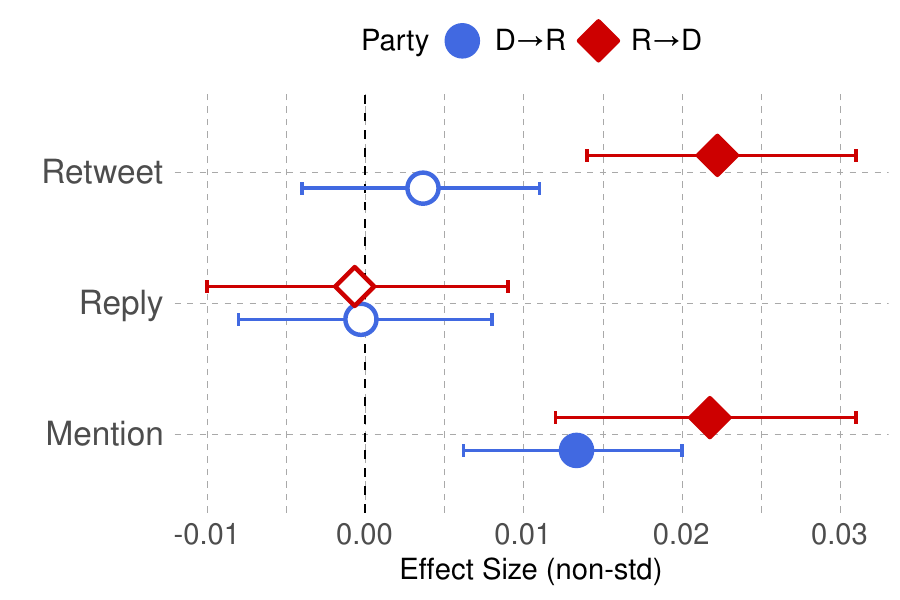}
        \caption{Non-standardized effect sizes of \b{engagement} on future rate of cross-cutting interactions (RQ2).}
        \label{fig:rq3_ns}
    \end{subfigure}

    \vspace{1em}

    \begin{subfigure}
        \centering
        \includegraphics[width=0.75\linewidth]{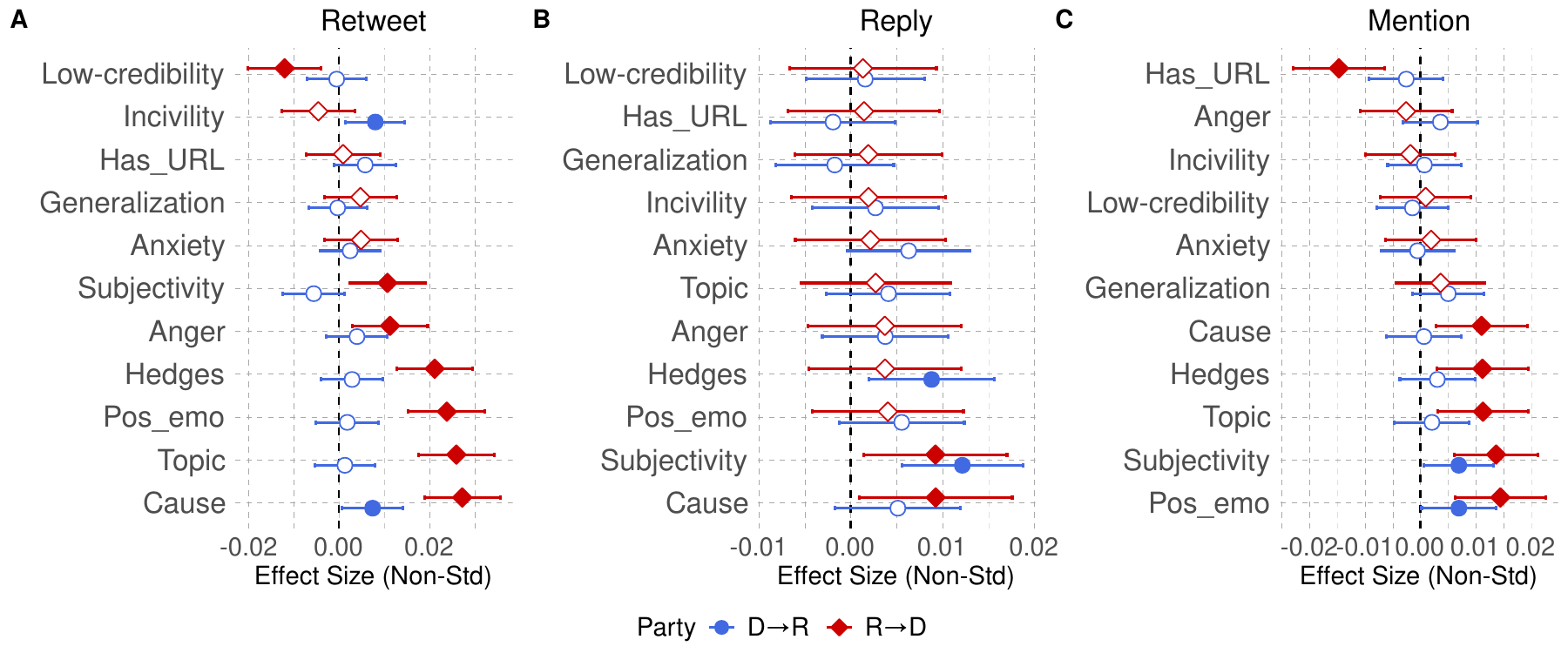}
        \caption{Non-standardized effect sizes of \b{engagement} on rhetoric and style of future cross-cutting interactions (RQ3).}
        \label{fig:rq4_ns}
    \end{subfigure}

\end{figure*}

\ab{\paragraph{Non-standardized Regression Results.}
Figures~\ref{fig:rq1_ns}, \ref{fig:rq2_ns}, \ref{fig:rq3_ns} and \ref{fig:rq4_ns} show the non-standardized effect sizes for RQ1a, 1b, 2, and 3 respectively. The non-standardized effect sizes (multiplying the coefficients by 100) can be interpreted as percentage changes in the outcome variable.}

\begin{table*}[ht]

\footnotesize
\setlength{\tabcolsep}{2pt}
\renewcommand{\arraystretch}{1}
\centering


\caption{
Examples of content characteristics related to the visibility of cross-cutting posts between Republicans and Democrats (\RBox\RtoD) and Democrats and Republicans (\DBox\DtoR) are provided. Features associated with higher visibility are marked with (+), while those linked to lower visibility are marked with (-). \emph{\b{Note: This table is illustrative rather than exhaustive and highlights representative content characteristics that are most strongly associated with visibility differences (RQ1b).}}
}\label{tab:rq2_example} 

\begin{tabular}{ccp{13.5cm} c}
\toprule
\textbf{Interaction} & \textbf{Feature} & \textbf{Examples} & No. \\ 

\midrule

\multirow{10}{*}{Retweet} & \multirow{2}{*}{Abortion} &  \raisebox{0.7ex}{\colorbox{BrickRed}{\hspace{0.1em}\vspace{0.5ex}}} \textcolor{BrickRed}{(-)} RT @xx: Seriously, though: for most of the religious right the "but Judges..." argument boils down to abortion and LGBT rights.  & \circled{1}\\
                              &                               & \raisebox{0.7ex}{\colorbox{RoyalBlue}{\hspace{0.1em}\vspace{0.5ex}}} RT @xx: I asked @xx three times this afternoon if he thinks something like Texas's 6-week abortion ban should be  & \circled{2}\\
                              
                              \cmidrule{3-3}
                              
                              &   \multirow{2}{*}{Gun}                            & \raisebox{0.7ex}{\colorbox{BrickRed}{\hspace{0.1em}\vspace{0.5ex}}} \textcolor{BrickRed}{(-)} RT @xx: In Texas you have to show strict ID to vote but not to carry a gun  & \circled{3}\\
                              &                               & \raisebox{0.7ex}{\colorbox{RoyalBlue}{\hspace{0.1em}\vspace{0.5ex}}} \textcolor{RoyalBlue}{(+)} RT @xx: Mass shootings and more heartbreak for families that lose loved ones to gun violence is a crisis in America.  & \circled{4}\\
                              
                              \cmidrule{3-3}
                              
                              &   \multirow{2}{*}{BLM}                            & \raisebox{0.7ex}{\colorbox{BrickRed}{\hspace{0.1em}\vspace{0.5ex}}} RT @xx: President @xx has ordered the FBI to expedite its investigation into the death of George Floyd.  & \circled{5}\\
                              &                       & \raisebox{0.7ex}{\colorbox{RoyalBlue}{\hspace{0.1em}\vspace{0.5ex}}} \textcolor{RoyalBlue}{(+)} RT @xx: George Floyd appears to have been the victim of murder. A close review of the video can lead one to no other conclusion  & \circled{6}\\
                              \cmidrule{3-3}
                              
                              &    \multirow{2}{*}{Incivility}                           & \raisebox{0.7ex}{\colorbox{BrickRed}{\hspace{0.1em}\vspace{0.5ex}}} \textcolor{BrickRed}{(-)} \ahb{RT @xx: Who is this white dude and why the fuck did he just ask this sista to give him examples of racism?}  & \circled{7}\\
                              &                              & \raisebox{0.7ex}{\colorbox{RoyalBlue}{\hspace{0.1em}\vspace{0.5ex}}} RT @xx: Donald Trump is the stupidest  & \circled{8}\\
                              \cmidrule{3-3}
                              
                              &   \multirow{2}{*}{Anger}                            & \raisebox{0.7ex}{\colorbox{BrickRed}{\hspace{0.1em}\vspace{0.5ex}}} RT @xx: We have the best military and the best intelligence anywhere in the world. If Americans are threatened, we are prepared..  & \circled{9}\\
                              &                             & \raisebox{0.7ex}{\colorbox{RoyalBlue}{\hspace{0.1em}\vspace{0.5ex}}} \textcolor{RoyalBlue}{(+)} RT @xx: This is testament the importance  of All Americans participating in the Democratic process. BIPOC cmty has fought for years.. & \circled{10}\\

\midrule
\multirow{10}{*}{Reply} & \multirow{2}{*}{Abortion} & \raisebox{0.7ex}{\colorbox{BrickRed}{\hspace{0.1em}\vspace{0.5ex}}} \textcolor{BrickRed}{(-)} .. Christians uphold Biblical values and honor scripture. How can you support \#abortion and \#homosexuality as a democrat and also be a sincere \#Christian..  & \circled{11}\\
                              &                               & \raisebox{0.7ex}{\colorbox{RoyalBlue}{\hspace{0.1em}\vspace{0.5ex}}} .. I support the right to an abortion, as does the Kansas constitution as interpreted by the KS Supreme Court .. I say reproductive freedoms because there is more to it than just abortion services & \circled{12}\\
                              \cmidrule{3-3}
                              
                              &   \multirow{2}{*}{Capitol}                            & \raisebox{0.7ex}{\colorbox{BrickRed}{\hspace{0.1em}\vspace{0.5ex}}} .. Today, men, women, young people, and children peacefully protested at their US Capitol. Only a small few broke the law. \#StopTheSteal  & \circled{13}\\
                              &                               & \raisebox{0.7ex}{\colorbox{RoyalBlue}{\hspace{0.1em}\vspace{0.5ex}}} \textcolor{RoyalBlue}{(-)} Weird, youd think a Patriot and veteran would have stopped people from attacking the nations Capitol right instead of running away ..  & \circled{14}\\
                              \cmidrule{3-3}
                              
                              &   \multirow{2}{*}{Incivility}                            & \raisebox{0.7ex}{\colorbox{BrickRed}{\hspace{0.1em}\vspace{0.5ex}}} \textcolor{BrickRed}{(+)} .. I have an issue with stupid rules that require mask wearing when no one is present .. so if they choose to have dumb rules that's their right. It's also my right to criticize them  & \circled{15}\\
                              &                       & \raisebox{0.7ex}{\colorbox{RoyalBlue}{\hspace{0.1em}\vspace{0.5ex}}} .. Would be threaten Dr Fauci- trumpsters, militia types - all those who misunderstand the meaning of freedom; the greater good Morons  & \circled{16}\\
                              \cmidrule{3-3}
                              
                              &    \multirow{2}{*}{Anger}                           & \raisebox{0.7ex}{\colorbox{BrickRed}{\hspace{0.1em}\vspace{0.5ex}}} \textcolor{BrickRed}{(+)} .. \ahb{by potential criminals \& untrained adults are you referencing Hunter Biden? Because surprisingly you haven't appeared to tweet about his lies on a federal form to obtain a gun which was later left in a public trash can}  & \circled{17}\\
                              &                              & \raisebox{0.7ex}{\colorbox{RoyalBlue}{\hspace{0.1em}\vspace{0.5ex}}} .. Maybe yall should consider not spreading lies and conspiracy theories then you might get to play with grownup toys \#GOPSeditiousTraitors \#Copolitics  & \circled{18}\\
                              \cmidrule{3-3}
                              
                              &   \multirow{2}{*}{Anxiety}                            & \raisebox{0.7ex}{\colorbox{BrickRed}{\hspace{0.1em}\vspace{0.5ex}}} \textcolor{BrickRed}{(+)} .. Current Missouri law outlaws the mere possession of brass knuckles, which is exactly the kind of thing many of us are worried some would do with guns. I don't see how your analogy applies  & \circled{19}\\
                              &                             & \raisebox{0.7ex}{\colorbox{RoyalBlue}{\hspace{0.1em}\vspace{0.5ex}}} Donald Trump is a danger to our country. He's unhinged and I personally fear what harm he has planned for Inauguration Day. We need to feel safe before we can heal \#TrumpIsDANGEROUS & \circled{20}\\

\midrule

\multirow{10}{*}{Mention} & \multirow{2}{*}{Abortion} & \raisebox{0.7ex}{\colorbox{BrickRed}{\hspace{0.1em}\vspace{0.5ex}}} \textcolor{BrickRed}{(-)} Even if the Texas abortion law is bad politics, it has inarguably already saved lives .. I'd rather good policies that save lives than good politics that save Republicans..  & \circled{21}\\
                              &                               & \raisebox{0.7ex}{\colorbox{RoyalBlue}{\hspace{0.1em}\vspace{0.5ex}}} .. People should make their own decisions about abortion without involving the government  & \circled{22}\\
                              \cmidrule{3-3}
                              
                              &   \multirow{2}{*}{Rights}                            & \raisebox{0.7ex}{\colorbox{BrickRed}{\hspace{0.1em}\vspace{0.5ex}}} .. Blaming a president or vice president? .. Demodupes keep getting angry that they're getting their rights trampled on by Democrat politicians, but refuse to vote out the culprits.  & \circled{23}\\
                              &                               & \raisebox{0.7ex}{\colorbox{RoyalBlue}{\hspace{0.1em}\vspace{0.5ex}}} \textcolor{RoyalBlue}{(+)} .. ending taxpayer subsidies for a foreign government that commits human rights abuses against millions of people is just so far beyond..  & \circled{24}\\
                              \cmidrule{3-3}
                              
                              &   \multirow{2}{*}{Gun}                            & \raisebox{0.7ex}{\colorbox{BrickRed}{\hspace{0.1em}\vspace{0.5ex}}} Don't let progressive liberal @xx take away your constitutional right to protect your loved ones..Gun sales in major swing states up nearly 80\% this year: Will it have any bearing on election outcome.. & \circled{25}\\
                              &                       & \raisebox{0.7ex}{\colorbox{RoyalBlue}{\hspace{0.1em}\vspace{0.5ex}}} \textcolor{RoyalBlue}{(+)} .. I am not embarrassed and will not apologize for receiving support from families and survivors of gun violence. This happened to my community and we are still dealing with the consequences..  & \circled{26}\\
                              \cmidrule{3-3}
                              
                              &    \multirow{2}{*}{Incivility}                           & \raisebox{0.7ex}{\colorbox{BrickRed}{\hspace{0.1em}\vspace{0.5ex}}} \textcolor{BrickRed}{(-)} .. Leftists live in an alternate reality which I refer to as leftist lunacy land where the Dems won the civil war. Revisionist history as they try to rewrite all history in their favor.. &  \circled{27}\\
                              &                              & \raisebox{0.7ex}{\colorbox{RoyalBlue}{\hspace{0.1em}\vspace{0.5ex}}} \textcolor{RoyalBlue}{(-)} .. Talk about Trump; my God the world is ending\! Trumps a demagogue, ignorant, idiotic; egotistical. GOP cares only about power, not the truth\! & \circled{28} \\
                              \cmidrule{3-3}
                              
                              &   \multirow{2}{*}{Generalization}                            & \raisebox{0.7ex}{\colorbox{BrickRed}{\hspace{0.1em}\vspace{0.5ex}}} \textcolor{BrickRed}{(-)} .. \#Patriots - stand tall and keep praying, tell all your friends. It will be fair, legal and legitimate or it will be nothing at all.. &  \circled{29}\\
                              &                             & \raisebox{0.7ex}{\colorbox{RoyalBlue}{\hspace{0.1em}\vspace{0.5ex}}} \textcolor{RoyalBlue}{(+)} .. the last thing Donald Trump needs in the world is this job President. Lets help him out and \#vote \#BidenHarris \#November3rd & \circled{30}\\

                              \cmidrule{3-3}

                            
                            
                            
                            

\bottomrule
\end{tabular}
\end{table*}
\begin{table*}[ht]
\footnotesize
\setlength{\tabcolsep}{2pt}
\renewcommand{\arraystretch}{1}
\centering
\captionsetup{list=no}
\caption[]{\textbf{Table~\ref{tab:rq2_example} (continued).} Examples of content characteristics related to the visibility of cross-cutting posts.}
\begin{tabular}{ccp{13.5cm} c}
\toprule
\textbf{Interaction} & \textbf{Feature} & \textbf{Examples} & No. \\ 
\midrule

\multirow{8}{*}{Mention} & \multirow{2}{*}{Causality} 
                            & \raisebox{0.7ex}{\colorbox{BrickRed}{\hspace{0.1em}\vspace{0.5ex}}} 
                            \textcolor{BrickRed}{(-)} .. \b{Nearly three dozen people died in Ontario because coronavirus policies delayed their heart surgeries..}  
                            & \circled{31}\\
                            & 
                            & \raisebox{0.7ex}{\colorbox{RoyalBlue}{\hspace{0.1em}\vspace{0.5ex}}} 
                            .. \b{They helped plan and bring terror and war to us and many other nations. Just sayin. Not like Taliban Afghanistan was just minding its own business and we went there..} 
                            & \circled{32}\\
                            
                            \cmidrule{3-3}
                            
                            & \multirow{2}{*}{Subjectivity} 
                            & \raisebox{0.7ex}{\colorbox{BrickRed}{\hspace{0.1em}\vspace{0.5ex}}} 
                            \textcolor{BrickRed}{(+)} .. \b{is the most liberal Republican representative Wyoming has EVER sent to D.C. Embarrassing!!! XX needs to resign.} 
                            & \circled{33}\\
                            & 
                            & \raisebox{0.7ex}{\colorbox{RoyalBlue}{\hspace{0.1em}\vspace{0.5ex}}} 
                            \textcolor{RoyalBlue}{(+)} .. \b{This could have something to do with our Top Ten Ranking in New Cases of COVID.} 
                            & \circled{34}\\
                            
                            \cmidrule{3-3}
                            
                            & \multirow{2}{*}{Hedges} 
                            & \raisebox{0.7ex}{\colorbox{BrickRed}{\hspace{0.1em}\vspace{0.5ex}}} 
                            \textcolor{BrickRed}{(-)} .. \b{Pressure mounts on XX to identify funders, organizers of violent riots | Just The News..} 
                            & \circled{35}\\
                            & 
                            & \raisebox{0.7ex}{\colorbox{RoyalBlue}{\hspace{0.1em}\vspace{0.5ex}}} 
                            .. \b{I hope XX gets a season in this year, but if even Jose Canseco says you're a 'fool', you gotta wonder if your plan makes sense..} 
                            & \circled{36}\\

\bottomrule
\end{tabular}
\end{table*}


\begin{table*}[t]
\centering
\setlength{\tabcolsep}{6pt}
\renewcommand{\arraystretch}{1.25}
\caption{\b{Prevalence of posting styles and representative examples by party. Percentages are computed over all posts containing at least one interaction (reply, mention, or retweet) in the estimable sample. Examples are illustrative and anonymized.}}
\label{tab:style_prevalence_examples}
\begin{tabularx}{\textwidth}{l c X X}
\toprule
\textbf{Posting style} & \textbf{\%} & \textbf{Republican example} & \textbf{Democrat example} \\
\midrule
Subjectivity & \b{52.5} & \textit{\b{I'm game if we can also require ID and confirm the actual holder of said ballot is actually the person casting the vote and that they are eligible to vote! I think we could all meet in the middle on this! Great opinion piece below.. }} & \textit{\b{..California can now sanction MediCal health care plans that don't comply with screening and testing the children most at risk of lead poisoning..}} \\
Causality & \b{35.1} & \textit{\b{..Why don't you look up the context of Ms. King's quote? It was made in reference to McCarthy's remarks about CRT; institutional racism. When I quoted Dr. King it was SOLELY in reference to the assertion that one is inherently racist based on their skin color. A big difference}} & \textit{\b{PFAS disposal is really just another step in the contamination cycle. Feeding the Waste Cycle: How PFAS Disposal Perpetuates Contamination..}} \\
Hedges & \b{37.1} & \textit{\b{..'I feel very nervous' about Biden's chances against Trump..}} & \textit{\b{..I wish we all realized that such training is not about blame or finger point though perhaps  uncomfortable at times, same as students often are as they learn, but we ask them to keep learning. If they can, we can. Hope we can inspire one another.}} \\
Generalization & \b{17.3} & \textit{\b{We must focus on STEM and rid ourselves of CRT..China Rises as Worlds STEM Leader While American Schools Place Diversity First..}} & \textit{\b{Once again, XX reports the truth about NY's inability to execute. This time it's administering COVID vaccinations..}} \\
Positive emotion & \b{57.6} & \textit{\b{Poll: Voters Favor Trump over Biden in Hypothetical 2024 Matchup..}} & \textit{\b{While XX is focused on bringing students safely back into the classroom, there are families who feel safer with at-home learning options. We will urge XX to make sure schools have the resources to expand online learning options for students.}} \\
Anger & \b{25.1} & \textit{\b{VICIOUS: Grown Woman Assaults 12-Year-Old Boy in Denver Over Pro-Trump Yard Sign..}} & \textit{\b{Ignorance now rules America. Not the simple, if somewhat innocent ignorance that comes from an absence of knowledge, but a malicious ignorance forged in the arrogance of refusing to think hard about an issue, to engage language in the pursuit of justice..}} \\
Anxiety & \b{20.3} & \textit{\b{While XX frantically repeats lies about the admins handling of the border, federal authorities are releasing thousands of Covid positive illegal aliens into the U.S..}} & \textit{\b{..And it was really critical here. Things will not be the same for large and small districts. Local as possible would have offered the least disruption.}} \\

\bottomrule
\end{tabularx}

\vspace{2pt}
\footnotesize{\textbf{Notes.} Posting styles are binary indicators (present/absent) as defined in Section~4.1. Examples are shortened and de-identified; they are intended to illustrate what each construct captures.}
\end{table*}



\end{document}